\documentclass[11pt,a4paper]{article}
\usepackage{amsfonts}
\usepackage{amssymb}
\usepackage[centertags]{amsmath}
\usepackage{graphicx}
\usepackage{booktabs}
\usepackage{theorem}
\usepackage{array}
\usepackage{ulem}
\usepackage[small,bf]{caption}
\usepackage{cite}
\usepackage{color}
\usepackage{colordvi}

\allowdisplaybreaks[1]

\usepackage{a4wide}
\numberwithin{equation}{section}

\newcommand{\be}{\begin{equation}}
\newcommand{\ee}{\end{equation}}
\newcommand{\beq}{\begin{equation}}
\newcommand{\eeq}{\end{equation}}
\newcommand{\bea}{\begin{eqnarray}}
\newcommand{\eea}{\end{eqnarray}}

\newcommand{\lv}{\langle}
\newcommand{\rv}{\rangle}

\DeclareMathOperator{\ci}{\text{i}}

\makeatletter

\newcommand{\Rmnum}[1]{\expandafter\@slowromancap\romannumeral #1@}
\makeatother

\begin{document}

\begin{titlepage}

\vspace*{-15mm}
\begin{flushright}
MPP-2013-13\\
SISSA 04/2013/FISI 
\end{flushright}
\vspace*{1.4cm}

\begin{center}
{
\bf\LARGE
Spontaneous CP violation in 
$\boldsymbol{A_4 \times SU(5)}$
with Constrained Sequential Dominance 2
}
\\[8mm]
Stefan~Antusch$^{\star}$
\footnote{E-mail: \texttt{stefan.antusch@unibas.ch}},
Stephen~F.~King$^{\dagger}$
\footnote{E-mail: \texttt{king@soton.ac.uk}},
Martin~Spinrath$^{\ddag}$
\footnote{E-mail: \texttt{spinrath@sissa.it}},
\\[1mm]
\end{center}
\vspace*{0.50cm}
\centerline{$^{\star}$ \it
 Department of Physics, University of Basel,}
\centerline{\it
Klingelbergstr.~82, CH-4056 Basel, Switzerland}
\vspace*{0.2cm}
\centerline{$^{\star}$ \it
Max-Planck-Institut f\"ur Physik (Werner-Heisenberg-Institut),}
\centerline{\it
F\"ohringer Ring 6, D-80805 M\"unchen, Germany}
\vspace*{0.2cm}
\centerline{$^{\dagger}$ \it
School of Physics and Astronomy, University of Southampton,}
\centerline{\it
SO17 1BJ Southampton, United Kingdom }
\vspace*{0.2cm}
\centerline{$^{\ddag}$ \it
SISSA/ISAS and INFN,}
\centerline{\it
Via Bonomea 265, I-34136 Trieste, Italy }
\vspace*{1.20cm}

\begin{abstract}
\noindent
We revisit a two right-handed neutrino model with two
texture zeros, namely an indirect model based on
$A_4$ with the recently proposed new type of constrained
sequential dominance (CSD2), involving vacuum alignments
along the $(0,1,-1)^T$ and  $(1,0,2)^T$ directions in
flavour space, which are proportional to the neutrino
Dirac mass matrix columns. In this paper we construct a
renormalizable and unified indirect $A_4 \times SU(5)$
model along these lines and show that, with spontaneous
CP violation and a suitable vacuum alignment of the phases,
the charged lepton corrections lead to a reactor angle in
good agreement with results from Daya Bay and RENO. The
model predicts a right-angled unitarity triangle in the
quark sector and a Dirac CP violating oscillation phase
in the lepton sector of $\delta \approx 130^\circ$,
while providing a good fit to all quark and lepton masses
and mixing angles.
\end{abstract}

\end{titlepage}

\setcounter{footnote}{0}

\section{Introduction}
The lepton mixing angles have the distinctive feature that the atmospheric
angle $\theta_{23}$ and the solar angle $\theta_{12}$, are both 
rather large \cite{pdg}.  Direct evidence for the reactor angle $\theta_{13}$ was
first provided by T2K, MINOS and Double
Chooz~\cite{Abe:2011sj,Adamson:2011qu,Abe:2011fz}. 
Subsequently Daya Bay~\cite{DayaBay}, RENO~\cite{RENO}, and Double
Chooz~\cite{DCt13} Collaborations have measured $\sin^2(2\theta_{13})$:
\begin{eqnarray}
\label{t13}
\begin{array}{cc}
\text{Daya Bay: } & \sin^2(2\theta_{13})=0.089\pm0.011 \text{(stat.)}\pm0.005
\text{(syst.)}\ ,\\
\text{RENO: }    & \sin^2(2\theta_{13})= 0.113\pm0.013\text{(stat.)}\pm0.019
\text{(syst.)\ ,}\\
\text{Double Chooz: }& \sin^2(2\theta_{13})=0.109\pm0.030\text{(stat.)}\pm0.025
\text{(syst.)}\ .\\
\end{array}
\end{eqnarray}

This rules out the hypothesis of exact tri-bimaximal (TB)
mixing~\cite{Harrison:2002er}, and many alternative proposals have
recently been put forward \cite{flurry}, although there are relatively few examples
which also include unification \cite{Antusch:2011qg, Marzocca:2011dh, GUTs, Meroni:2012ty}.
For example, an attractive scheme based on trimaximal (TM) mixing remains viable 
\cite{Haba:2006dz}, sometimes referred to as TM$_2$ mixing since it maintains the second column of the
TB mixing matrix and hence preserves the solar mixing angle prediction
$\sin\theta_{12} \approx 1/\sqrt{3}$. However there is another variation of TM
mixing which also preserves this good solar mixing angle prediction by
maintaining the first column of the TB matrix, namely TM$_1$ mixing
\cite{Lam:2006wm}. 

Although there were models of TM$_2$ mixing which can account for the smallness 
of the reactor angle \cite{King:2011zj}, the first model in the literature
for TM$_1$ mixing, which also fixed the value of the reactor angle, 
was proposed in \cite{Antusch:2011ic}. The model discussed in \cite{Antusch:2011ic} 
was actually representative of a
general strategy for obtaining TM$_1$ mixing using sequential dominance (SD)
\cite{King:1998jw} and vacuum alignment. The strategy of combining SD with
vacuum alignment is familiar from the constrained sequential dominance (CSD)
approach to TB mixing \cite{King:2005bj} where a neutrino mass hierarchy is
assumed and the dominant and subdominant flavons responsible for the
atmospheric and solar neutrino masses are aligned in the directions of the 
third and second columns of the TB mixing matrix, namely 
$\lv\phi_1^{\nu}\rv \propto  (0,1,-1)^T$ and 
$\lv\phi_2^{\nu}\rv\propto (1,1,1)^T$.
The new idea was to maintain the usual vacuum alignment for the dominant
flavon,  $\lv\phi_1^{\nu}\rv\propto (0,1,-1)^T$ as in CSD, but to replace the
effect of the subdominant flavon vacuum alignment by a different one, namely
either $\lv\phi_{120}\rv\propto (1,2,0)^T$ or $\lv\phi_{102}\rv\propto (1,0,2)^T$,
where such alignments may be naturally achieved from the standard ones using
orthogonality arguments.

We referred to this new approach as CSD2 
\footnote{It is interesting to compare the predictions of CSD2 to another alternative to CSD that 
has been proposed to account for a reactor angle called partially constrained
sequential dominance (PCSD) \cite{King:2009qt}. PCSD involves a vacuum
misalignment of the dominant flavon alignment to $(\varepsilon ,1,-1)^T$, with a 
subdominant flavon alignment $(1,1,1)^T$, leading to tri-bimaximal-reactor
(TBR) mixing \cite{King:2009qt} in which only the reactor angle is switched
on, while the atmospheric and solar angles retain their TB values. }
and showed
that it leads to TM$_1$ mixing and a reactor angle which, at leading
order, is predicted to be proportional to the ratio of the solar to the
atmospheric neutrino masses, $\theta_{13} = \frac{\sqrt{2}}{3} \,
\frac{m^\nu_2}{m^\nu_3}$. The model was proposed before 
the results from Daya Bay and RENO, and the prediction turned out to be rather too small
compared to the results in Eq.\ \eqref{t13}. More generally it has been shown that any type I seesaw
model with two right-handed neutrinos and two texture zeros in the neutrino Yukawa matrix
(as in Occam's razor)
is not compatible with the experimental data for the case of a normal neutrino mass hierarchy
\cite{Harigaya:2012bw}. However this conclusion ignores the effect of charged lepton corrections,
and so an ``Occam's razor''  model which includes such corrections may become viable.

In the present paper we construct a fully renormalisable
unified $A_4 \times SU(5)$ model in which the neutrino sector
satisfies the CSD2 conditions, and show that, with spontaneous CP violation and 
a suitable vacuum alignment of the phases, the charged lepton corrections can correct the 
reactor angle, bringing it into agreement with results from Daya Bay and RENO. 
We shall use here similar techniques as in \cite{Antusch:2011sx}, where spontaneous CP violation 
with flavon phases determined by the vacuum alignment was discussed for the first time, in order
to ensure that the charged lepton mixing angle correction
(typically about $\sim 3^\circ$)
adds constructively to the $\theta_{13}^{\nu}$ angle from the neutrino sector 
(typically about $\sim 5^\circ - 6^\circ$)
leading to $\theta_{13} \sim 8^\circ - 9^\circ$, within the range of the measured value from Daya Bay and RENO.
In fact the present model is more ambitious, since it describes all quark and lepton masses and mixing angles, including predictions for all the CP violating phases.

We demonstrate the viability of the model by performing a global fit to the charged lepton masses and the quark masses and mixing parameters. For the neutrino mixing angles we make a parameter scan and find very good agreement with the experimental data. 
We emphasise that the present $A_4 \times SU(5)$ model represents one of the first unified ``indirect'' family symmetry models in the literature that has been constructed to date that is consistent with all experimental data on quark and lepton mass and mixing parameters
where ``indirect'' simply means that the family symmetry is completely broken by the vacuum 
alignment.\footnote{In fact the only other example of a unified indirect model with a realistic reactor angle that 
we are aware of is the last paper in \cite{GUTs} based on Pati-Salam unification, however that model 
predicts an atmospheric angle in second octant.}
For a review see \cite{King:2013eh}.

We emphasise that the idea of spontaneous CP violation has a long history \cite{Branco:2011zb}. However, in explicit flavour
models using this idea only the positions of the phases in the mass matrices was predicted,
but not the phases of the flavon fields themselves (see e.g.\ \cite{Ross:2004qn}).
Spontaneous CP violation with calculable flavon phases 
from vacuum alignment was first discussed in \cite{Antusch:2011sx} and demonstrated in example models based on $A_4$ and 
$S_4$.
In this paper we shall use a similar approach where the
$A_4$ model is formulated in the real $SO(3)$ basis (see e.g.\ \cite{King:2006np}) and where we only consider the real representations $\mathbf{1}$ and $\mathbf{3}$. In such a framework, one can either use a ``simple'' CP symmetry under which the components of the scalar fields transform trivially as 
$\phi_i \rightarrow \phi_i^{*}$, or a ``generalised'' CP symmetry which intertwines CP with $A_4$ (see e.g.\ \cite{Feruglio:2012cw} and 
references therein). In the latter case, in our basis, the triplet fields would transform as $\phi_i \rightarrow U_3 \phi_i^{*}$, where $U_3$ interchanges the second and third component. When complex $\mathbf{1}'$ and $\mathbf{1}''$ representations are used in a model, the $U_3$ transformation then takes care of the fact that under CP the two complex singlets are interchanged with each other. However, as already mentioned above, in our model this will make no difference. CP symmetry leads to real coupling constants in a suitable field basis (after ``unphysical'' phases have been absorbed by field redefinitions). CP is subsequently spontaneously broken by the flavon vacuum alignment, which is controlled by additional Abelian symmetries $\mathbb{Z}_3$ and $\mathbb{Z}_4$, resulting in calculable complex flavon phases as in  \cite{Antusch:2011sx}.

The layout of the rest of the paper is as follows: 
in the next section we discuss the general strategy we will adopt in our model.
After a brief review of CSD2 we discuss charged lepton sector corrections to
TM$_1$ mixing before we describe the method which we use to fix the flavon vevs.
In section \ref{Sec:Model} we describe our model, the field content and symmetries
and the resulting Yukawa and mass matrices. The justification for the chosen vacuum
alignment including phases is given in section \ref{Sec:Flavon}. In the subsequent sections
we comment on the Higgs mass and then we give the numerical results from our global fit
and scans. In section \ref{Sec:Summary} we summarize and conclude and in the appendix
we define our notations and conventions and give the messenger sector of our model.

\section{The strategy}
\label{Sec:Strategy}

Let us now describe our general idea in somewhat more detail, before we present an explicit GUT model example in the next section. As outlined in the introduction, we are combining three ingredients which finally result in a highly predictive unified flavour model. These ingredients are:
\begin{itemize}
\item CSD2 for the neutrino mixing angles $\theta_{ij}^\nu$,
\item charged lepton mixing contributions as they are typical in GUTs,
\item spontaneous CP violation with aligned phases.
\end{itemize}
We now briefly describe these three concepts and the resulting new class of models.

\subsubsection*{CSD2 in the neutrino sector} 
In models with CSD2 \cite{Antusch:2011ic}, the neutrino mass matrix is dominated by two right-handed neutrinos with mass matrix $M_R={\rm diag}(M_A, M_B)$ and couplings to the lepton doublets $A = (0,a,-a)^T$ and $B = (b,0,2b)^T$ \footnote{$B = (b,2b,0)^T$ was also considered in  \cite{Antusch:2011ic}, but here we shall not consider it further.} such that the neutrino Yukawa matrix takes the form $Y_{\nu} = (A, B)$, in left-right convention. A summary of the used conventions is given in the Appendix.

After the seesaw mechanism is implemented,
CSD2 leads to the following light effective 
neutrino Majorana mass matrix:
\begin{equation}
M_\nu = m_a \begin{pmatrix} 0 & 0 & 0 \\ 0 & 1 & -1 \\ 0 & -1 & 1 \end{pmatrix} + m_b \begin{pmatrix} 1 & 0 & 2 \\ 0 & 0 & 0 \\ 2 & 0 & 4  \end{pmatrix} = m_a \begin{pmatrix} \epsilon \, \text{e}^{\ci \alpha} & 0 & 2 \epsilon \, \text{e}^{\ci \alpha} \\ 0 & 1 & -1 \\ 2 \epsilon \, \text{e}^{\ci \alpha} & -1 & 1 + 4 \epsilon \, \text{e}^{\ci \alpha} \end{pmatrix},
\label{eq:numassmatrix}
\end{equation}
where $m_a = \frac{v_u^2  a^2}{M_A}$, $m_b = \frac{v_u^2  b^2}{M_B}$, and where 
$\alpha$ is the relative phase difference between $m_a$ and $m_b$. We define $\epsilon = |m_b|/|m_a|$, 
and assume $ \epsilon \ll 1$ leading to a normal mass hierarchy
in accordance with SD. 
As discussed in Appendix \ref{App:Conventions} we use
here different conventions than in the original CSD2 paper \cite{Antusch:2011ic} which
are more convenient in the context of $SU(5)$ GUTs.

Only three parameters, e.g.\ $m_a$, $\epsilon$ and $\alpha$, govern the neutrino masses and mixing parameters. For the mixing parameters, the predicted values are, to leading order in $\epsilon$ (from \cite{Antusch:2011ic} with adapted conventions\footnote{Compared to the notation of \cite{Antusch:2011ic}, we have changed, for instance, $\alpha \rightarrow - \alpha$.}):
\begin{align}
 s_{23}^\nu &\approx \frac{1}{\sqrt{2}} - \frac{\epsilon}{\sqrt{2}} \cos \alpha \;, & \delta_{13}^\nu &\approx \pi - \alpha + \epsilon \frac{5}{2} \sin \alpha \;, \\
 s_{13}^\nu &\approx \frac{\epsilon}{\sqrt{2}} \;, & \alpha_2 &\approx  - \alpha + 2 \epsilon \sin \alpha \;, \\
 s_{12}^\nu &\approx \frac{1}{\sqrt{3}} \;.
\end{align}
The mixing scheme resulting from CSD2 can be identified as trimaximal mixing of type 1 (i.e.\ TM$_1$ \cite{Lam:2006wm}) but with a predicted value of the neutrino 1-3 mixing, $\theta^\nu_{13} = \frac{\sqrt{2}}{3}\frac{m^\nu_2}{m^\nu_3} \sim 5^\circ - 6^\circ$. With neutrino mass $m^\nu_1=0$, only one Majorana CP phase is physical. Without charged lepton corrections, $\delta_{13}^\nu$ would be identical with the leptonic Dirac CP phase $\delta$. Let us also note that CSD2 predicts a deviation of $\theta_{23}^\nu$ from $45^\circ$, depending on the phase $\alpha$. 

\subsubsection*{Charged lepton mixing contribution in GUTs} 
In GUT models the charged lepton Yukawa matrix is generically non-diagonal in the flavour basis, due to the close link between the charged lepton and the down-type quark Yukawa matrices, which typically provides the main origin of the flavour mixing in the quark sector. With the Cabibbo angle $\theta_C$ being the largest mixing in the quark sector, the mixing in $Y_e$ is often dominated by a 1-2 mixing $\theta^e_{12}$ as well, such that the relevant part of (the hierarchical matrix) $Y_e$ can be written as
\begin{equation}
 Y_e \approx \begin{pmatrix}
       0 & c \, \text{e}^{\ci \beta} & 0 \\
       * & d & 0 \\
       0 & * & * \end{pmatrix} \;,
\end{equation}
where $c$, $d$, and $\beta$ are real and where the entries marked by a '$*$' are not relevant for our discussion here. With $c \ll d$ one can read off to leading order the values for the complex  1-2 mixing angle (for more details see also Appendix \ref{App:Conventions}) that are
\begin{equation}
 \theta_{12}^e \approx \left| \frac{c}{d} \right| \text{ and } \delta^e_{12} = \begin{cases} -\beta & \text{for } c/d >0 \\ - \beta + \pi & \text{for } c/d < 0 \end{cases} \;.
\end{equation}
Since we will have $c/d < 0$ in our example GUT model in the next section, let us consider this case also in the following discussion.  

In explicit GUT models, $\theta^e_{12}$ is typically related to the Cabibbo angle by group theoretical Clebsch factors from GUT breaking, as has been discussed recently, e.g.\ in \cite{Antusch:2011qg,Marzocca:2011dh}. In many GUT models, in particular in those where the muon and the strange quark mass at the GUT scale is predicted by such a Clebsch factor as $m_\mu/m_s = 3$  \cite{Georgi:1979df}, but also if the Yukawa matrices $Y_e$ (and $Y_d$) are (nearly) symmetric with a zero in the (0,0)-element \cite{Antusch:2011qg}, $\theta_{12}^e$ is predicted as
\begin{equation}
 \theta_{12}^e \approx \frac{ \theta_C}{3} \;.
\end{equation}
In the example GUT model in the next section we will see explicitly how such a prediction arises in an $SU(5)$ GUT.

The leptonic mixing parameters, defined via $U_{\text{PMNS}} = U_{e} U_{\nu}^\dagger$, are a combination of the mixing from the neutrino and the charged lepton sectors. Making use of the fact that, to leading order, $\theta_{23}^e = \theta_{13}^e = \delta_{12}^\nu = \delta_{23}^\nu = 0$, and using the CSD2 expressions from above for the neutrino sector, and general formulae for the charged lepton mixing contributions of \cite{Antusch:2005kw,King:2005bj,Antusch:2008yc}
\begin{align}
s_{23} \text{e}^{- \ci \delta_{23}} &= s_{23}^\nu \text{e}^{- \ci \delta_{23}^\nu} - \theta_{23}^e c_{23}^\nu \text{e}^{- \ci \delta_{23}^e}   \;, \\
s_{13} \text{e}^{- \ci \delta_{13}} &= \theta_{13}^\nu \text{e}^{- \ci \delta_{13}^\nu} - \theta_{12}^e s_{23}^\nu \text{e}^{- \ci (\delta_{23}^\nu + \delta_{12}^e)}  \;, \\
s_{12} \text{e}^{- \ci \delta_{12}} &= s_{12}^\nu \text{e}^{- \ci \delta_{12}^\nu} - \theta_{12}^e c_{23}^\nu c_{12}^\nu \text{e}^{- \ci \delta_{12}^e} \;,
\end{align}
we obtain (up to $\mathcal{O}(\epsilon)$)
\begin{align}
 \theta_{23} &\approx 45^\circ - \epsilon \cos \alpha  \;, \\
\theta_{13} &\approx  \frac{\epsilon}{\sqrt{2}}  -  \cos(\beta - \alpha)  \,\frac{\theta_{12}^e}{\sqrt{2}}  \;, \\
\theta_{12} &\approx 35.3^\circ + \cos \beta\, \frac{\theta_{12}^e}{\sqrt{2} } \;,
\end{align}
where $\epsilon \approx  \frac{2}{3}\frac{m^\nu_2}{m^\nu_3} \approx 8.4^\circ$. When the phases $\alpha$ and $\beta$ are fixed by the vacuum alignment, and when also $\theta_{12}^e$ is predicted from the GUT structure, as both will be the case in our model, all three mixing angles and also the CP phases $\delta$ and $\alpha_2$, are predicted. Thus, the resulting models of this type can be highly predictive.

We would like to note here already that in the explicit GUT model in the next section, we will construct a vacuum alignment such that $\alpha = \pi / 3$, leading to\footnote{We note that the choice $\alpha = \pi/3$ is motivated by the current data which favours $\theta_{23}$ in the first octant. On the other hand, one can in principle also construct other models with different values of $\alpha$, and there are also other options for $\beta$ and $\theta_{12}^e$, which may lead to interesting alternative models. In this sense, the strategy described here leads to a whole new class of possible models.}
 \begin{equation}
\theta_{23} \approx 45^\circ - \frac{\epsilon}{2}   \approx 41^\circ \:, 
\end{equation}
close to the best fit value for the normal hierarchy case
from global fits to the neutrino data \cite{Fogli:2012ua}. The alignment of $\beta$ will satisfy $\beta =  \alpha + \pi$, such that the neutrino and charged lepton contributions to $\theta_{13}$ simply add up, leading to (with $\theta_{12}^e = \theta_C/3$)
 \begin{equation}
 \theta_{13} \approx \frac{\epsilon}{\sqrt{2}} + \frac{ \theta_C}{3 \sqrt{2}}  \approx 8^\circ - 9^\circ\;,
\end{equation}
in agreement with the recent measurements. 
With these values of $\alpha$ and $\beta$, it also turns out that $\theta_{12}$ is predicted somewhat smaller than $35^\circ$, namely
\begin{equation}
\theta_{12} \sim 33^\circ \;.
\end{equation}
This value of $\theta_{12}$ could be distinguished from the tribimaximal value by a future reactor experiment with $\sim 60$ km baseline \cite{Minakata:2004jt}.

\subsubsection*{Spontaneous CP violation with aligned phases} 
Finally, the third ingredient is spontaneous CP violation with aligned phases of the flavon vevs, using the method proposed in \cite{Antusch:2011sx}. To give a brief summary of this method, let us note that phase alignment can very simply be achieved using discrete symmetries when the flavon vevs effectively depend on one parameter, i.e. when the direction of the vevs is given by the form of the potential. 
This remains true even in the presence of ``generalised'' CP transformations as long as these CP transformations fix the phases of the involved coupling constants. 
Working example models with $A_4$ and $S_4$ family symmetry can be found in \cite{Antusch:2011sx}. Note that $S_4$ is in agreement only with ``simple'' CP, while the ``generalised'' CP transformation for $A_4$ interchanges the complex singlet representations
\cite{Feruglio:2012cw}. In both cases all the coupling constants are forced to be real in a suitable field basis.

To illustrate the phase alignment, let us consider a case with a flavon field $\xi$ which is a singlet under the family symmetry and singly charged under a $\mathbb{Z}_n$ shaping symmetry (with $n\ge2$). Then typical terms in the flavon superpotential, which ``drive'' the flavon vev non-zero, have the form
\begin{equation}\label{eq:flavonpotentialZn}
P \left( \frac{\xi^n}{\Lambda^{n-2}} \mp M^2 \right).
\end{equation} 
The field $P$ is the so-called ``driving superfield'', meaning that the $F$-term $|F_P|^2$ generates the potential for $\xi$ which enforces a non-zero vev. $\Lambda$ is the (real and positive) suppression scale of the effective operator, and $M$ here is simply a (real) mass scale. 
From the potential for $\xi$,
\begin{equation}
|F_P|^2 =  \left| \frac{\xi^n}{\Lambda^{n-2}} \mp M^2 \right|^2 ,
\end{equation} 
the vev of $\xi$ has to satisfy
\begin{equation}\label{eq:flavonvevsdiscrete}
\xi^n = \pm \,\Lambda^{n-2}  M^2\;.
\end{equation} 
Since the right side of the equation is real, we obtain that
\begin{equation}\label{eq:phaseswithZn}
\arg(\langle \xi \rangle) =   \left\{ \begin{array}{ll}
\frac{2 \pi}{n}q \;,\quad q = 1, \dots , n & \mbox{\vphantom{$\frac{f}{f}$} for ``$-$'' in Eq.~(\ref{eq:flavonpotentialZn}),}\\
\frac{2 \pi}{n} q +\frac{\pi}{n} \;,\quad q = 1, \dots , n & \mbox{\vphantom{$\frac{f}{f}$} for ``$+$'' in Eq.~(\ref{eq:flavonpotentialZn}).}
\end{array}
\right.
\end{equation} 
For example, with a $\mathbb{Z}_3$ shaping symmetry and a ``$+$'' in Eq.~(\ref{eq:flavonpotentialZn}), only multiples of $\pi/3$ are allowed for $\arg( \langle \xi \rangle )$. 
We will use this method for the relevant flavons to constrain their phases. In the ground state, one of the vacua (with a fixed phase) is selected, which finally determines also the two phases $\alpha$ and $\beta$ relevant for the predictions in the lepton sector. 

Furthermore, we note that we will also use the phase alignment to generate the CP violation in the quark sector, predicting a right-angled unitarity triangle, which is in excellent agreement with the present data (making use of the quark phase sum rule from \cite{Antusch:2009hq}). 

We now turn to an explicit GUT model, where the above described strategy is applied.

\section{The model}
\label{Sec:Model}
In the following we will construct an  $A_4 \times SU(5)$ model with CSD2 \cite{Antusch:2011ic} in the neutrino sector. The model follows the strategy described in the previous section, such that the charged lepton mixing contribution to $\theta_{13}$ adds up constructively with the  1-3 mixing in the neutrino sector to $\theta_{13} \sim 8^\circ - 9^\circ$, with the phases fixed by the ``discrete vacuum alignment'' mechanism \cite{Antusch:2011sx}.

\begin{table}
\centering
\begin{tabular}{c cccccc cccccc}
\toprule
& $T_1$ & $T_2$ & $T_3$ & $F$ & $N_1$ & $N_2$ & $H_5$ & $\bar H_5$ & $H_{45}$ & $\bar H_{45}$ & $H_{24}$ & $S$  \\
\midrule $SU(5)$ & $\mathbf{10}$ & $\mathbf{10}$ & $\mathbf{10}$ & $\mathbf{\bar 5}$
& $\mathbf{1}$ & $\mathbf{1}$ & $\mathbf{5}$ & $\mathbf{\bar 5}$ & $\mathbf{45}$ & $\mathbf{\overline{45}}$ & $\mathbf{24}$  & $\mathbf{1}$ \\
$A_4$ & $\mathbf{1}$ & $\mathbf{1}$ & $\mathbf{1}$ & $\mathbf{3}$ & $\mathbf{1}$ & $\mathbf{1}$ & $\mathbf{1}$
& $\mathbf{1}$ & $\mathbf{1}$ & $\mathbf{1}$ & $\mathbf{1}$ & $\mathbf{1}$  \\
$U(1)_R$ & 1 & 1 & 1 & 1 & 1 & 1 & 0 & 0 & 0 & 0 & 0 & 2 \\ 
\midrule
$\mathbb{Z}_4$ & 3 & 3 & 3 & 0 & 0 & 2 & 2 & 0 & 0 & 2 & 1 & 2 \\ 
$\mathbb{Z}_4$ & 3 & 3 & 3 & 0 & 2 & 2 & 2 & 0 & 2 & 0 & 1 & 2 \\ 
$\mathbb{Z}_3$ & 1 & 2 & 0 & 0 & 1 & 2 & 0 & 0 & 0 & 0 & 2 & 0 \\ 
$\mathbb{Z}_3$ & 1 & 1 & 0 & 0 & 2 & 0 & 0 & 0 & 1 & 2 & 2 & 0 \\ 
$\mathbb{Z}_3$ & 0 & 2 & 2 & 1 & 0 & 2 & 2 & 0 & 1 & 1 & 2 & 1 \\ 
$\mathbb{Z}_3$ & 0 & 0 & 0 & 2 & 0 & 0 & 0 & 0 & 1 & 2 & 1 & 0 \\
\bottomrule
\end{tabular}
\caption{\label{Tab:Matter+HiggsFields}
The matter and Higgs fields in our model and their quantum numbers.}
\end{table}

\begin{table}
\centering
\begin{tabular}{c ccccc ccccc cccc}
\toprule
 & $SU(5)$ & $A_4$ & $U(1)_R$ & $\mathbb{Z}_4$ & $\mathbb{Z}_4$ & $\mathbb{Z}_3$ & $\mathbb{Z}_3$ & $\mathbb{Z}_3$ & $\mathbb{Z}_3$ \\ 
 \midrule 
$\phi_{102} $ &  $\mathbf{1}$ &  $\mathbf{3}$ &  0  &  0  &  0  &  1  &  0  &  1  &  1\\
$\phi_{23} $  &  $\mathbf{1}$ &  $\mathbf{3}$ &  0  &  2  &  0  &  2  &  1  &  0  &  1\\
$\phi_{1} $ &  $\mathbf{1}$ &  $\mathbf{3}$ &  0  &  1  &  3  &  1  &  0  &  0  &  1\\
$\phi_{2} $ &  $\mathbf{1}$ &  $\mathbf{3}$ &  0  &  0  &  3  &  0  &  0  &  0  &  0\\
$\phi_{3} $ &  $\mathbf{1}$ &  $\mathbf{3}$ &  0  &  1  &  1  &  0  &  0  &  0  &  1\\
$\phi_{111} $ &  $\mathbf{1}$ &  $\mathbf{3}$ &  0  &  3  &  3  &  0  &  0  &  0  &  0\\
$\phi_{211} $ &  $\mathbf{1}$ &  $\mathbf{3}$ &  0  &  0  &  0  &  2  &  1  &  1  &  0\\
\midrule
$\xi_{u} $ &  $\mathbf{1}$ &  $\mathbf{1}$ &  0  &  0  &  0  &  0  &  2  &  1  &  0\\
$\xi_{1} $ &  $\mathbf{1}$ &  $\mathbf{1}$ &  0  &  0  &  0  &  1  &  2  &  0  &  0\\
$\xi_{2} $ &  $\mathbf{1}$ &  $\mathbf{1}$ &  0  &  0  &  0  &  2  &  0  &  2  &  0\\
\midrule
$\theta_{2} $ &  $\mathbf{1}$ &  $\mathbf{1}$ &  0  &  0  &  1  &  0  &  0  &  0  &  0\\
$\theta_{102} $ &  $\mathbf{1}$ &  $\mathbf{1}$ &  0  &  0  &  0  &  1  &  0  &  0  &  2\\
\midrule
$\rho_{111} $ &  $\mathbf{1}$ &  $\mathbf{1}$ &  0  &  3  &  3  &  0  &  0  &  0  &  0\\
$\tilde \rho_{111} $  &  $\mathbf{1}$ &  $\mathbf{1}$ &  0  &  3  &  3  &  0  &  0  &  0  &  0\\
$\rho_{23} $  &  $\mathbf{1}$ &  $\mathbf{1}$ &  0  &  0  &  0  &  2  &  1  &  0  &  1\\
$\rho_{102} $ &  $\mathbf{1}$ &  $\mathbf{1}$ &  0  &  0  &  0  &  1  &  0  &  1  &  1\\
\bottomrule
\end{tabular}
\caption{\label{Tab:FlavonFields}
The flavon field content of our model.}
\end{table}

The matter and the Higgs
sector of the model is summarised in Table~\ref{Tab:Matter+HiggsFields}
while the required flavons are shown in Table~\ref{Tab:FlavonFields}.
The superpotential after integrating out the heavy
messenger fields, see Appendix~\ref{App:Messenger},
and suppressing order one coefficients reads 
\begin{align}
 \mathcal{W}_N &= \xi_1 N_1^2 + \xi_2 N_2^2 \;,\\
 \mathcal{W}_\nu &= \frac{1}{\Lambda} (H_5 F) (\phi_{23} N_1) + \frac{1}{\Lambda} (H_5 F) (\phi_{102} N_2) \;,\\
 \mathcal{W}_d &= \frac{1}{\Lambda^3} \theta_2 \bar H_5 F (T_1 \phi_2) H_{24} + \frac{1}{\Lambda^3} \theta_{102} \bar H_5 F (T_2 \phi_{102}) H_{24}
  + \frac{1}{\Lambda^2} F (T_2 \phi_{23}) \bar H_{45} H_{24} + \frac{1}{\Lambda} \bar H_5 F (T_3 \phi_3) \;,\\
  \mathcal{W}_u &= \frac{1}{\Lambda^2} T_1^2 H_5 \xi_u \xi_1 + \frac{1}{\Lambda^2} T_1 T_2  H_5 \xi_u^2 + \frac{1}{\Lambda^2} T_2^2 H_5 \xi_1^2 + \frac{1}{\Lambda} T_2 T_3 H_5 \xi_1 + T_3^2 H_5 \;,
\end{align}
where $\Lambda$ denotes the messenger scale.
The flavon potential, which gives rise to the vevs of the fields $\phi_i$, $\xi_i$ and $\theta_i$ will be discussed separately in the next section. Note that the flavons of type $\phi$ which enter the Yukawa couplings will be aligned with real vevs while the flavons of type $\theta$ and $\xi$ will generally acquire complex vevs with precisely determined phases.
The above superpotential gives rise to the flavour structures in the neutrino sector, in the down-type quark and charged lepton sectors, and in the up-type quark sector.

{\bf Neutrino sector:} From the flavon potential, to be discussed in the next section, the two triplet flavons entering the neutrino Yukawa sector are aligned along the directions
\begin{align} \label{eq:FlavonDirectionsNu}
 \lv \phi_{23} \rv  \sim \begin{pmatrix} 0 \\ 1 \\ -1 \end{pmatrix}, \ \ \ \   
 \lv \phi_{102} \rv \sim \begin{pmatrix} 1 \\ 0 \\  2 \end{pmatrix} \;, \quad
 \end{align}
where both alignments are real. Inserting the above vacuum alignments,
the real vev $\lv \xi_1 \rv$ and the vev $\lv \xi_2 \rv$ with a phase of $-\pi / 3$
into the superpotential leads to a
Dirac Yukawa matrix and a right-handed
heavy Majorana mass matrix of the form:
\begin{equation}
 Y_\nu = \begin{pmatrix}
	 0 &   b \\
	 a &   0 \\
	-a & 2 b
         \end{pmatrix} \text{ and } \;\;
M_R = \begin{pmatrix}
	M_A & 0 \\
	0 & M_B
      \end{pmatrix} .
\end{equation}
where $M_A$,  $a$ and $b$ are real
and $M_B$ has a complex phase of $- \pi/3$.
The (type-I) seesaw
formula leads to a simple effective light neutrino
mass matrix of the form given in eq.~\eqref{eq:numassmatrix}
where the relative phase
difference $\alpha$ between $m_a$ and $m_b$ is
now fixed to be $\pi/3$.
This form of $M_\nu$ gives 
$\theta^\nu_{13} \sim 5^\circ - 6^\circ$ for the 1-3 mixing in the neutrino sector,
which will finally add up with the charged lepton
mixing contribution. 

{\bf Down-type quark and charged lepton sector:} Turning to the down quark and charged lepton sector, two further triplet flavons enter: 
\begin{align} \label{eq:FlavonDirectionsDown}
 \lv \phi_{2} \rv   &\sim \begin{pmatrix} 0 \\ 1 \\  0 \end{pmatrix} \;, \quad
 \lv \phi_{3} \rv   \sim \begin{pmatrix} 0 \\ 0 \\  1 \end{pmatrix} \;,
  \end{align}
where $\lv \phi_{2} \rv $ is aligned to be real.  The 
phase of $\lv \phi_{3} \rv$ will turn out to be unphysical.
Furthermore the singlet $\theta_2$ with a phase of $\pi/2$
and the singlet $\theta_{102}$ with a phase of $4 \pi/3$
enters.
Plugging in the vevs of the flavon fields leads to the following structure of the
Yukawa matrices (in left-right convention) for the down-type quarks and
charged leptons:
\begin{align} \label{eq:YdYe}
 Y_d = \begin{pmatrix}
	0 & \ci \epsilon_2 & 0 \\
	\bar\omega \epsilon_{102} & \epsilon_{23} & 2 \bar\omega  \epsilon_{102} - \epsilon_{23} \\
	0 & 0 & \epsilon_3 
       \end{pmatrix} \text{ and }
 Y_e =  \begin{pmatrix}
	0 & -3/2 \bar\omega  \epsilon_{102} & 0 \\
	- 3/2 \ci \epsilon_2 &  9/2 \, \epsilon_{23} & 0 \\
	0 & \left(-3 \bar\omega  \epsilon_{102} - 9/2 \, \epsilon_{23} \right) & \epsilon_3
       \end{pmatrix} \;,
\end{align}
where $\bar\omega = \text{e}^{4 \pi \ci/3}$, cf.\ section \ref{Sec:Strategy}.
The $\epsilon_i$ are proportional to the order one couplings which
we have not written down explicitly and possible Higgs mixing angles.
For the sake of simplicity we only show here the proportionality to the
dimensionful quantities
\begin{align}
 \epsilon_2 \sim \frac{v_{24}}{\Lambda^3} |\lv \theta_2 \rv \lv \phi_2 \rv| \text{, }
 \epsilon_{102} \sim \frac{v_{24}}{\Lambda^3} |\lv \theta_{102} \rv \lv \phi_{102} \rv| \text{, }
 \epsilon_{23} \sim \frac{v_{24}}{\Lambda^2} |\lv \phi_{23} \rv| \text{, }
 \epsilon_{3} \sim \frac{1}{\Lambda} |\lv \phi_{3} \rv| \text{,}
\end{align}
where $v_{24}$ is the vev of $H_{24}$.
We also note that we do not use the common
Georgi-Jarlskog relation $m_\mu / m_s = 3$ \cite{Georgi:1979df} at the GUT 
scale but rather $m_\mu / m_s = 9/2$  \cite{Antusch:2009gu,Antusch:2008tf}. 
The reason for this is that recent lattice results, see, e.g.\
\cite{Juttner:2011jg} suggest a much smaller error
for the strange quark mass than the PDG quotes.
And since we are in the small $\tan \beta$ regime
and no large SUSY threshold corrections can correct
the second generation GUT scale Yukawa coupling
ratios we have to use the more realistic relation mentioned above.
Explicitly, from the vevs of $H_{24}$ and $\bar H_5$ we get
a relative factor of $-3/2$ for $\epsilon_2$ and
$\epsilon_{102}$ and the $9/2$ from $H_{24}$ and
$\bar H_{45}$. For the third generation we use $b-\tau$ Yukawa
unification which is possible for small $\tan \beta$
due to the large RGE effects induced by the top mass.

For the 1-2 mixing in the charged lepton sector, we nevertheless obain $\theta_{12}^e \approx \theta_C/3$, where $\theta_C
\approx 0.23$ is the Cabibbo angle. The corresponding phase $\delta_{12}^e$ is chosen
(see section~\ref{Sec:Strategy} and appendix~\ref{App:Conventions} for conventions), such that the charged
lepton mixing angle correction $\theta_{12}^e$ is in
phase with the neutrino reactor angle  $\theta_{13}^{\nu}$
 and the two angles add together constructively to yield
the physical reactor angle  $\theta_{13}$. 

{\bf Up-type quark sector:} 
Finally the up-type quark sector only involves singlet
flavons with real vevs and gives a
real symmetric Yukawa matrix of the form,
\begin{equation}
 Y_u = \begin{pmatrix}
	a_u & b_u & 0 \\
	b_u & c_u & d_u \\
	0 & d_u & e_u
       \end{pmatrix} \;,
\end{equation}
where the dependence on $\Lambda$ and the flavon vevs reads
\begin{equation}
a_u \sim \frac{|\lv \xi_u \rv \lv \xi_1 \rv|}{\Lambda^2} \text{, }
b_u \sim \frac{|\lv \xi_u \rv|^2}{\Lambda^2} \text{, }
c_u \sim \frac{|\lv \xi_1 \rv|^2}{\Lambda^2} \text{, }
d_u \sim \frac{|\lv \xi_u \rv|}{\Lambda} \text{.} 
\end{equation}
Note that $e_u$ is coming from a renormalisable coupling
and we have not explicitly written down all coefficients.
For instance, $\Lambda$ is only a simplified notation for the
various messenger masses as given in Appendix~\ref{App:Messenger},
and hence $a_u^2 \ll |b_u c_u|$ as in our numerical fit in
Section~\ref{sec:fit} is possible.
The zero texture in the quark sector means that we can successfully apply the quark phase sum rule of \cite{Antusch:2009hq} due to our choice of phases.

\section{The vacuum alignment}
\label{Sec:Flavon}

We have in total seven flavon fields which transform
as triplets under $A_4$, see Table~\ref{Tab:FlavonFields},
pointing in the following directions in flavour space,
\begin{align} \label{eq:FlavonDirections}
 \lv \phi_{1} \rv   \sim \begin{pmatrix} 1 \\ 0 \\  0 \end{pmatrix} \;, \quad
 \lv \phi_{2} \rv   &\sim \begin{pmatrix} 0 \\ 1 \\  0 \end{pmatrix} \;, \quad
 \lv \phi_{3} \rv   \sim \begin{pmatrix} 0 \\ 0 \\  1 \end{pmatrix} \;,  \quad \\
 \lv \phi_{211} \rv \sim \begin{pmatrix} -2 \\ 1 \\ 1 \end{pmatrix} \;, \quad
 \lv \phi_{111} \rv  & \sim \begin{pmatrix}  1 \\ 1 \\ 1 \end{pmatrix} \;, \quad
 \lv \phi_{23} \rv  \sim \begin{pmatrix} 0 \\ 1 \\ -1 \end{pmatrix} \;.
  \end{align}
Apart from $\lv \phi_{1} \rv $ and $\lv \phi_{3} \rv $, the vevs of the above listed flavons will be aligned real using the phase alignment mechanism proposed in \cite{Antusch:2011sx}. The phases of $\lv \phi_{1} \rv $ and $\lv \phi_{3} \rv $ have no physical implications
and hence will be set real for definiteness.
The first three vevs form a basis in flavour space, while the
second three alignments are proportional to 
the (real) columns of the
tri-bimaximal mixing matrix. In our model, instead of
$\phi_{111}$ (which is used in the CSD
\cite{King:1998jw, King:2005bj} models), we require the
following (real) alignment,
 \begin{align}
 \lv \phi_{102} \rv \sim \begin{pmatrix} 1 \\ 0 \\  2 \end{pmatrix} \;,
 \end{align}
in the neutrino sector, similar to a recently proposed flavon
alignment \cite{Antusch:2011ic} but with the phase
fixed as explicitly shown and discussed below.

The principal assumption of our model is that CP is
conserved above the flavour breaking scale, and is 
spontaneously broken by the CP violating phases of flavon fields. 
With this assumption we can not only reproduce the
correct mixing angles but can also make definite testable
predictions for the CP violating phases in the lepton sector.
In order to do this
we will fix the phases of the following flavon vevs to
\begin{align} \label{Eq:Phases}
 \alpha_{111} = 0 \;, \quad
 \alpha_{211} = 0 \;, \quad
 \alpha_{23}  = 0 \;, \quad
 \alpha_{2}  = 0 \;, \quad
 \alpha_{102}  = 0 \;, 
\end{align}
where $\alpha_i$ stands for the phase of $\lv \phi_i \rv$.
Furthermore we have some singlet flavons with non-vanishing
vevs of which some will have non-trivial phases. In this choice we
have also ignored possible signs which means that the phases
are fixed up to $\pm \pi$.
We can fix the phases by using appropriate $\mathbb{Z}_n$ shaping
symmetries as described in our previous paper \cite{Antusch:2011sx},
see also section~\ref{Sec:Strategy}.

\begin{table}
\centering
\begin{tabular}{c ccccc ccccc cccc}
\toprule
 & $SU(5)$ & $A_4$ & $U(1)_R$ & $\mathbb{Z}_4$ & $\mathbb{Z}_4$ & $\mathbb{Z}_3$ & $\mathbb{Z}_3$ & $\mathbb{Z}_3$ & $\mathbb{Z}_3$ \\ 
 \midrule 
$O_{1;2} $  &  $\mathbf{1}$ &  $\mathbf{1}$ &  2  &  3  &  2  &  2  &  0  &  0  &  2\\
$O_{1;3} $  &  $\mathbf{1}$ &  $\mathbf{1}$ &  2  &  2  &  0  &  2  &  0  &  0  &  1\\
$O_{2;3} $  &  $\mathbf{1}$ &  $\mathbf{1}$ &  2  &  3  &  0  &  0  &  0  &  0  &  2\\
$O_{111;211} $  &  $\mathbf{1}$ &  $\mathbf{1}$ &  2  &  1  &  1  &  1  &  2  &  2  &  0\\
$O_{111;23} $ &  $\mathbf{1}$ &  $\mathbf{1}$ &  2  &  3  &  1  &  1  &  2  &  0  &  2\\
$O_{23;211} $ &  $\mathbf{1}$ &  $\mathbf{1}$ &  2  &  2  &  0  &  2  &  1  &  2  &  2\\
$O_{2;102} $  &  $\mathbf{1}$ &  $\mathbf{1}$ &  2  &  0  &  1  &  2  &  0  &  2  &  2\\
$O_{211;102} $  &  $\mathbf{1}$ &  $\mathbf{1}$ &  2  &  0  &  0  &  0  &  2  &  1  &  2\\
$O_{1;23} $ &  $\mathbf{1}$ &  $\mathbf{1}$ &  2  &  1  &  1  &  0  &  2  &  0  &  1\\
\midrule
$A_{1} $ &  $\mathbf{1}$ &  $\mathbf{3}$ &  2  &  2  &  2  &  1  &  0  &  0  &  1\\
$A_{2} $ &  $\mathbf{1}$ &  $\mathbf{3}$ &  2  &  0  &  2  &  0  &  0  &  0  &  0\\
$A_{3} $ &  $\mathbf{1}$ &  $\mathbf{3}$ &  2  &  2  &  2  &  0  &  0  &  0  &  1\\
$A_{111} $ &  $\mathbf{1}$ &  $\mathbf{3}$ &  2  &  2  &  2  &  0  &  0  &  0  &  0\\
\midrule
$P$  &  $\mathbf{1}$ &  $\mathbf{1}$ &  2  &  0  &  0  &  0  &  0  &  0  &  0\\
\bottomrule
\end{tabular}
\caption{\label{Tab:DrivingFields}
The driving field content of our model. Note that we only
show here one $P$ field. Indeed one has to introduce as
many $P$ fields as operators to fix the phases of the
flavon fields. Since they will have all the same quantum
numbers they will mix and we can go to a basis where
the terms to fix the phase for each flavon is separated
from the others. This was discussed in the appendix
of \cite{Antusch:2011sx}.
}
\end{table}

The method can be understood easily for the $A_4$
singlet flavon vevs. Their superpotential reads
\begin{align}
 \mathcal{W} &= \frac{P}{\Lambda} (\xi_1^3 - M^3) + \frac{P}{\Lambda} (\xi_2^3 + M^3) + \frac{P}{\Lambda} (\xi_u^3 - M^3) \nonumber\\
 &+  \frac{P}{\Lambda^2} (\theta_2^4 - M^4) + \frac{P}{\Lambda} (\theta_{102}^3  - M^3) + \frac{P}{\Lambda} (\rho_{102}^3  - M^3)  + \frac{P}{\Lambda} (\rho_{23}^3  - M^3) \;,
\end{align}
where $M$ is a generic mass scale which we assume
to be positive. The list of the driving fields is given
in Table~\ref{Tab:DrivingFields}.
The $F$ terms for $P$ will then fix
the flavon vevs of the singlets up to a discrete choice.
Note that for the sake of simplicity we have only
introduced one $P$ field. Indeed, we need one
$P$ field for every singlet. Since they all have the
same quantum numbers they will mix and we
can go to a basis where all terms are disentangled
as in the equation above, see the appendix of
\cite{Antusch:2011sx}.
For the singlet flavons here we choose $\lv \xi_{1,u} \rv$
and $\lv \rho_{102,23} \rv$ to be real,
$\lv \theta_{2} \rv$ to be imaginary,
$\lv \theta_{102} \rv$ to have a phase of $4 \pi/3$
and $\lv \xi_2 \rv$ to have a phase of $- \pi/3$.

We come now back to the phases of the triplet flavon vevs
which can be fixed in the same way
after the direction in flavour space is fixed.
Note that the phases $\alpha_1$ and $\alpha_3$
are not fixed in our model. This is also not necessary. The
flavon $\phi_1$ does not couple to the matter sector
and hence its phase does not appear in the mass
matrices. It will only be used in orthogonality
relations where the phase of the vev does not matter.
The flavon $\phi_3$ couples nevertheless to the
matter sector. But as we have seen before it determines
the 3-3 element of the down-type quark and charged
lepton Yukawa matrix and its phase can be absorbed
in the right-handed fields such that this phase
renders unphysical.

In this section we will use an explicit notation for
the contraction of the $A_4$ indices. We use the
standard ``$SO(3)$ basis''
for which the singlet of ${\bf 3 \otimes 3}$ is given by
the $SO(3)$-type inner product '$\cdot$'. The two triplets of
${\bf 3 \otimes 3}$ are constructed from the usual
(antisymmetric) cross product '$\times$' and the
symmetric star product '$\star$' (see, for example,
\cite{King:2006np}).

We start with the alignment of the triplet flavons $\phi_i$, $i=1,2,3$,
which can be aligned via
\begin{equation}
 \mathcal{W} = A_i \cdot (\phi_i \star \phi_i) + O_{i;j} (\phi_i \cdot \phi_j) + \frac{P}{\Lambda^2} \left( (\phi_2 \cdot \phi_2)^2 - M^4 \right) \;.
\end{equation}
Solving the $F$-term conditions of $A_i$ aligns the
flavons in one of the three standard directions and the
$F$-term conditions of $O_{i;j}$ makes them orthogonal
to each other. By convention we
let them point in the directions as given in
eq.\ \eqref{eq:FlavonDirections}. For $\alpha_2$ we
choose the value $0$ ($\alpha_1$ and $\alpha_3$
remain undetermined). In Appendix \ref{App:Messenger}
we will discuss the messenger sector of our model.
After integrating out heavy messenger fields we end
up only with the effective operators written here and in the
following.

We now turn to the flavons $\phi_{23}$, $\phi_{111}$ and
$\phi_{211}$: For $\phi_{111}$ we use a slight
modification of the alignment in the recent $SU(5) \times T^{\prime}$
model \cite{Meroni:2012ty} without auxiliary flavons,
\begin{align}
\mathcal{W} &=
    A_{111} \cdot \left( \phi_{111} \star \phi_{111} + \phi_{111} \rho_{111}+ \phi_{111} \tilde\rho_{111}\right)
  + \frac{P}{\Lambda^2} \left( (\phi_{111} \cdot \phi_{111})^2 - M^4 \right) \nonumber\\
 & + \frac{P}{\Lambda^2} \left(\rho_{111}^4 + \rho_{111}^2 \tilde \rho_{111}^2 + \tilde \rho_{111}^4 - M^4 \right) \;.
\end{align}
It gives the desired alignment and $\lv \phi_{111} \rv$
can be chosen to be real.

Starting from this the other two alignments can be realised by
\begin{equation}
\begin{split}
 \mathcal{W} &= O_{1;23} (\phi_1 \cdot \phi_{23}) + O_{111;23} (\phi_{111} \cdot \phi_{23})
	+ O_{111;211} (\phi_{111} \cdot \phi_{211} ) + O_{23;211} (\phi_{23} \cdot \phi_{211} ) \\
	&\quad + \frac{P}{\Lambda} \left( ( \phi_{211} \star \phi_{211}) \cdot \phi_{211} - M^3 \right) + \frac{P}{\Lambda} \left( (\phi_{23} \cdot \phi_{23} ) \rho_{23} - M^3 \right) \;.
\end{split}
\end{equation}
The orthogonality gives the desired directions and
$\lv \phi_{211} \rv$ can be chosen to be real.
The phase of $\lv \phi_{23} \rv$ is a bit peculiar.
Above we have fixed $\lv \rho_{23} \rv$ to be real
and hence also $\lv \phi_{23} \rv$ can be chosen to
be real.
In the first operator the vev of $\phi_1$
enters again and independent of the phases
a $(0,1,-1)$ alignment is always orthogonal
to a $(1,0,0)$ alignment. 

Now we have everything together for the last missing
non-trivial alignment
\begin{equation}
 \mathcal{W} = O_{211;102} (\phi_{102} \cdot \phi_{211}) + O_{2;102} (\phi_{102} \cdot \phi_2) + \frac{P}{\Lambda} \left( (\phi_{102} \cdot \phi_{102}) \rho_{102} - M^3 \right) \;.
\end{equation}
The direction is again fixed by orthogonality conditions.
The vev of $\phi_{102}$ can be chosen to be real
(remember that also $\lv \rho_{102} \rv$ is real).

\section{The Higgs mass}

In our model we assume $b-\tau$ Yukawa coupling unification
at the GUT scale. This happens in the MSSM only for large
$\tan \beta$ via SUSY threshold corrections or small $\tan
\beta$ due to large RGE corrections by the top mass. We have
decided for the second solution such that we can also
neglect SUSY threshold corrections in our fit later on.

Nevertheless, the MSSM with small $\tan \beta$ prefers
very light Higgs masses which is in conflict with the
recent discovery of a Higgs-like particle with a mass
of about 126~GeV \cite{Higgs}.

A possible solution to this problem is given by the
NMSSM, for a review see \cite{Ellwanger:2009dp} where the Higgs can have
the right mass even for small $\tan \beta$. In fact
our symmetries forbid a $\mu$-term because the combinations
$H_5 \bar H_5$ and $H_{45} \bar H_{45}$ are charged
under the shaping symmetries. But we have checked
that we can add a singlet field $S$ which couples
simultaneously to this two combinations.
For convenience we have listed the field $S$
in Table~\ref{Tab:Matter+HiggsFields}.

An explicit $S^3$ term in the superpotential is
forbidden in the limit of unbroken $U(1)_R$ symmetry 
(i.e.\ before SUSY breaking) and by the shaping
symmetries but is needed to stabilize the Higgs potential
in the scale invariant NMSSM. But we note that there are
still various possibilities 
to stabilize the potential for $S$. This could be done, for instance, 
by introducing an additional $U(1)'$ gauge group where the
potential is stabilized by the $U(1)'$ $D$-terms.
For a description of this and references, see the
review article \cite{Ellwanger:2009dp}. We only note
that it is straightforward to introduce such a $U(1)'$
in our model by charging the Higgs and matter fields
appropriately which does not alter the flavour sector.
Alternatively, the $S^3$ term could be generated
non-perturbatively, breaking the shaping symmetries in
an $F$-theory framework, see, for instance,
\cite{Callaghan:2012rv}. We will not go here into more 
detail on this model building aspect and only like to note that 
our flavour model is compatible with some NMSSM variants 
and hence we can have a realistic Higgs mass.

\section{The fit and numerical results}\label{sec:fit}

Here we will present the results of a numerical $\chi^2$-fit
of the high energy parameters of the Yukawa matrices
to the low energy charged lepton and quark masses and
quark mixing parameters. Afterwards we will present
the predictions for neutrino masses and mixing.

\begin{figure}
\centering
\includegraphics[scale=1]{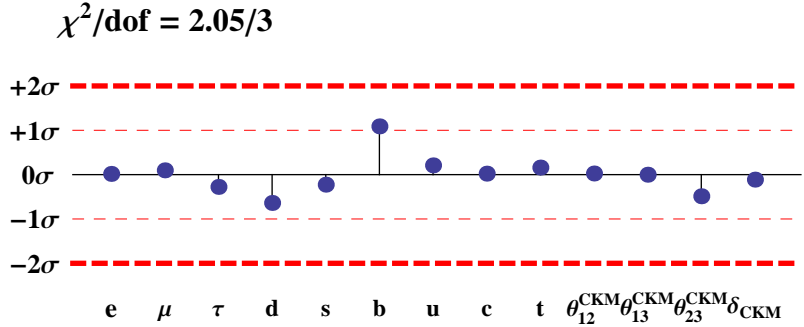}
\caption{Pictorial representation of the deviation
of our predictions from low energy experimental data
for the charged lepton Yukawa couplings and quark Yukawa
couplings and mixing parameters.  The deviations of the
charged lepton masses are given in~1\% while all other
deviations are given in units of standard deviations $\sigma$.
\label{Fig:FitResultsPlot} }
\end{figure}

\begin{table}
\begin{center}
\begin{tabular}{cc}
\toprule
Parameter & Value \\ \midrule
$a_u$ & $-3.01 \cdot 10^{-5}$ \\
$b_u$ & $-2.66 \cdot 10^{-4}$ \\
$c_u$ & $-2.57 \cdot 10^{-3}$ \\
$d_u$ & $ 3.09 \cdot 10^{-2}$\\
$e_u$ & $2.05$ \\
\midrule
$\epsilon_2$ & $-3.57 \cdot 10^{-5}$ \\
$\epsilon_{102}$ & $3.17 \cdot 10^{-5}$ \\
$\epsilon_{23}$ & $1.62 \cdot 10^{-4}$ \\
$\epsilon_3$ & $1.24 \cdot 10^{-2}$ \\
\midrule
$\tan \beta$ & $1.49$ \\
\bottomrule
\end{tabular}
\caption{Values of the effective parameters of the quark and
charged lepton Yukawa matrices and $\tan \beta$ for
$M_{\text{SUSY}} = 750$~GeV. The
numerical values are determined from a $\chi^2$-fit to experimental
data with a $\chi^2$ per degree of freedom of 2.05/3.
\label{Tab:Parameters}}
\end{center}
\end{table}

\begin{table}
\begin{center}
\begin{tabular}{cccc}
\toprule
Quantity (at $m_t(m_t)$)& Experiment & Model & Deviation \\ \midrule
$y_\tau$ in $10^{-2}$   & 1.00          & $1.00$ & $-0.277$ \\
$y_\mu$ in $10^{-4}$    & 5.89          & $5.89$ & $ 0.097$ \\
$y_e$ in $10^{-6}$  	& 2.79          & $2.79$ & $-0.016$ \\
\midrule
$y_b$ in $10^{-2}$      & $1.58 \pm 0.05$        & $1.64$ & $1.088$ \\
$y_s$ in $10^{-4}$  	& $2.99 \pm 0.86$        & $2.95$ & $-0.226$ \\
$y_d$ in $10^{-6}$      & $15.9^{+6.8}_{-6.6}$   & $11.7$ & $-0.639$ \\
\midrule
$y_t$                   & $0.936 \pm 0.016$      & $0.939$& $ 0.159$ \\
$y_c$ in $10^{-3}$      & $3.39 \pm 0.46$        & $3.40$ & $ 0.223$ \\
$y_u$ in $10^{-6}$      & $7.01^{+2.76}_{-2.30}$ & $7.59$ & $0.209$ \\
\midrule
$\theta_{12}^{\text{CKM}}$ & $0.2257^{+0.0009}_{-0.0010}$ & $0.2257$ & $ 0.026$ \\[0.3pc]
$\theta_{23}^{\text{CKM}}$ & $0.0415^{+0.0011}_{-0.0012}$ & $0.0409$ & $-0.488$ \\[0.1pc]
$\theta_{13}^{\text{CKM}}$ & $0.0036 \pm 0.0002$          & $0.0036$ & $-0.002$ \\[0.1pc]
$\delta_{\text{CKM}}$ & $1.2023^{+0.0786}_{-0.0431}$      & $1.1975$ & $-0.113$ \\
\bottomrule
\end{tabular}
\caption{Fit results for the quark Yukawa couplings and mixing and
the charged lepton Yukawa couplings at low energy compared to
experimental data. The values for the Yukawa couplings are extracted
from \cite{Xing:2007fb} and the CKM parameters from \cite{PDG10}.
Note that the experimental uncertainty on the charged lepton Yukawa
couplings are negligible small and we have assumed a relative
uncertainty of 1~\% for them. The $\chi^2$ per degree of freedom
is 2.05/3. A pictorial representation of the agreement between our
predictions and experiment can be found as well in
Fig.~\ref{Fig:FitResultsPlot}. \label{Tab:FitResults}}
\end{center}
\end{table}

For the RGE running of the Yukawa matrices we
have used the {\tt REAP} package \cite{Antusch:2005gp}
and calculated with it the masses and mixing angles
at low energies.
Note that we have used the RGEs of the MSSM. Possible
RGE effects due to including a variant of the NMSSM
are neglected. On the one hand we can expect this effect
to be flavour blind leading only to a rescaling
of the GUT scale parameters and on the
other hand, in the scale-invariant NMSSM for example, 
the RGE effects come from the coupling $\lambda$
which can be small \cite{Kowalska:2012gs}
although $\tan \beta$ given there is preferred to be
larger than 10.
For small $\tan \beta$ the coupling $\lambda$
has to be rather large to be in agreement with
recent Higgs data, see, e.g.~\cite{Cheng:2013fma}.
Furthermore, SUSY threshold corrections are negligibly small
due to the small $\tan \beta$ and hence are not
included in the fit.

For the charged lepton and quark masses
and their errors at the top scale $m_t(m_t)$
we have taken the values from \cite{Xing:2007fb}
and for the CKM parameters the PDG values \cite{PDG10}.
Note that the experimental errors for the
charged lepton masses are tiny and we have
estimated the theoretical uncertainty
from higher order effects to 1~\%, and 
we will assume this as their errors instead.

The Yukawa matrices depend on nine real parameters
(five from the up-type quarks and four from
the down-type quarks and charged leptons).
Furthermore we have included $\tan \beta$
as a free parameter in the fit. The unification
of the $b$ and the $\tau$ Yukawa coupling at the GUT scale
depends strongly on this parameter. On the contrary, 
the masses and mixing angles depend only very weakly
on the SUSY scale which we have therefore fixed to
$M_{\text{SUSY}} = 750$~GeV.

The fit results are summarised in Figure~\ref{Fig:FitResultsPlot}
and Tables~\ref{Tab:Parameters} and \ref{Tab:FitResults}.
We have fitted ten parameters to thirteen observables
with a $\chi^2$ of 2.05 and hence we can say that our
model describes the data very well.\footnote{We note that while we get an excellent fit 
for the quark masses themselves, as given in the PDG review, there is some tension with QCD results
which favour $y_s/y_d \approx 19$ \cite{Leutwyler:2000hx}, while our fit
yields $y_s/y_d = 25.3$. We remark that this tension is the same that one also gets with the more conventional
GJ relation instead of the Clebsch factors $9/2$ and $3/2$ used here, so it is not particular for our model. 
In our fit, we have not included $y_s/y_d$ as constraints, but we would like to note that future even more 
precise results on the quark masses, including lattice results, can provide powerful additional 
constraints on unified flavour models.}
Note that we followed here the strategy of our previous paper \cite{Antusch:2009hq}
where we have found that for Yukawa matrices with negligibly small 1-3 mixings
we find the correct value for the CKM phase and the Cabibbo angle $\theta_C$
with $\theta_{12}^d \approx \epsilon_2/\epsilon_{23} \lesssim \theta_C$
and $\theta_{12}^u \approx b_u/c_u \approx \theta_C/2$ if these two angles
have a relative phase difference of 90$^\circ$.

\begin{figure}
\centering
\includegraphics[scale=0.45]{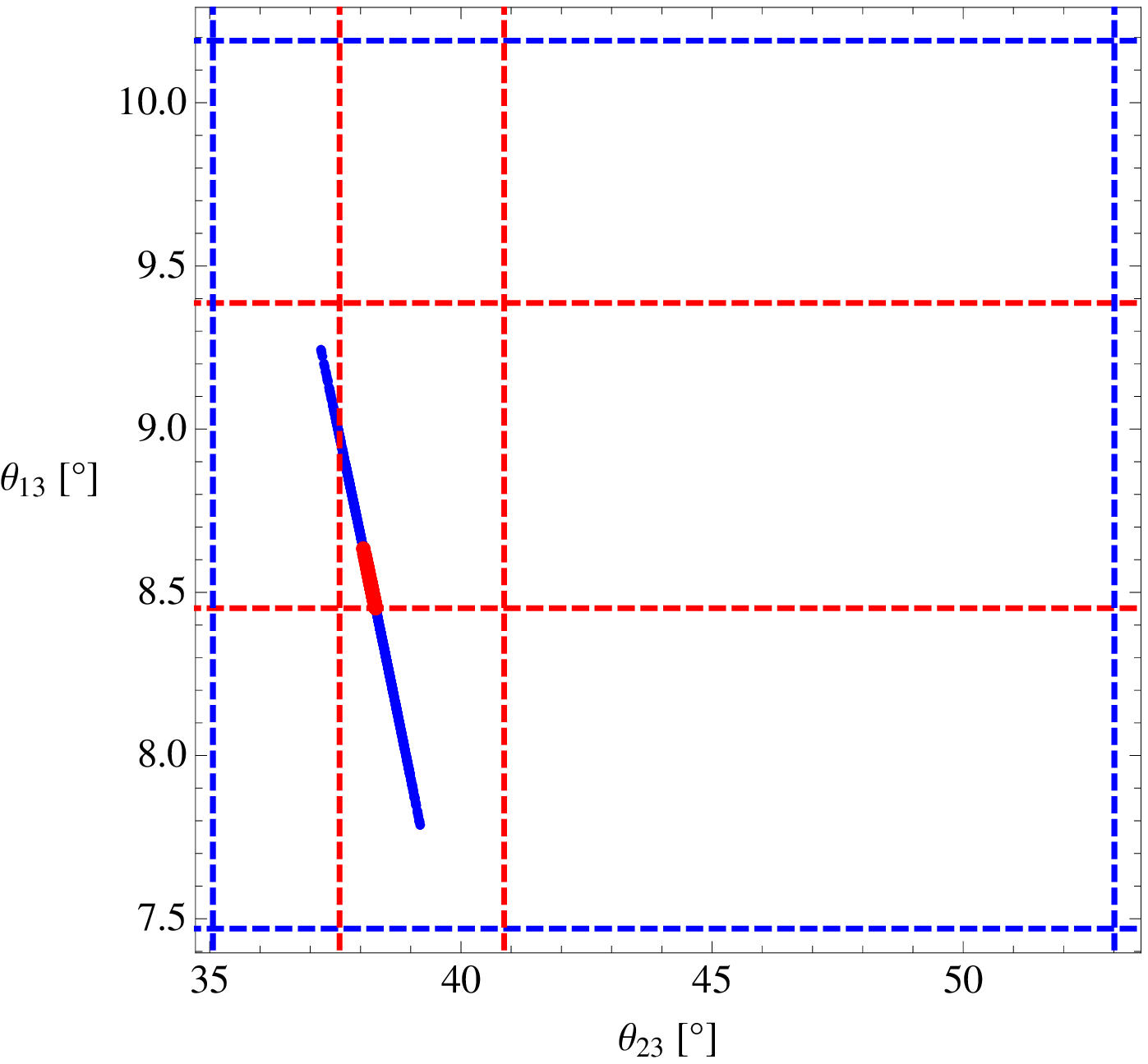}
\includegraphics[scale=0.45]{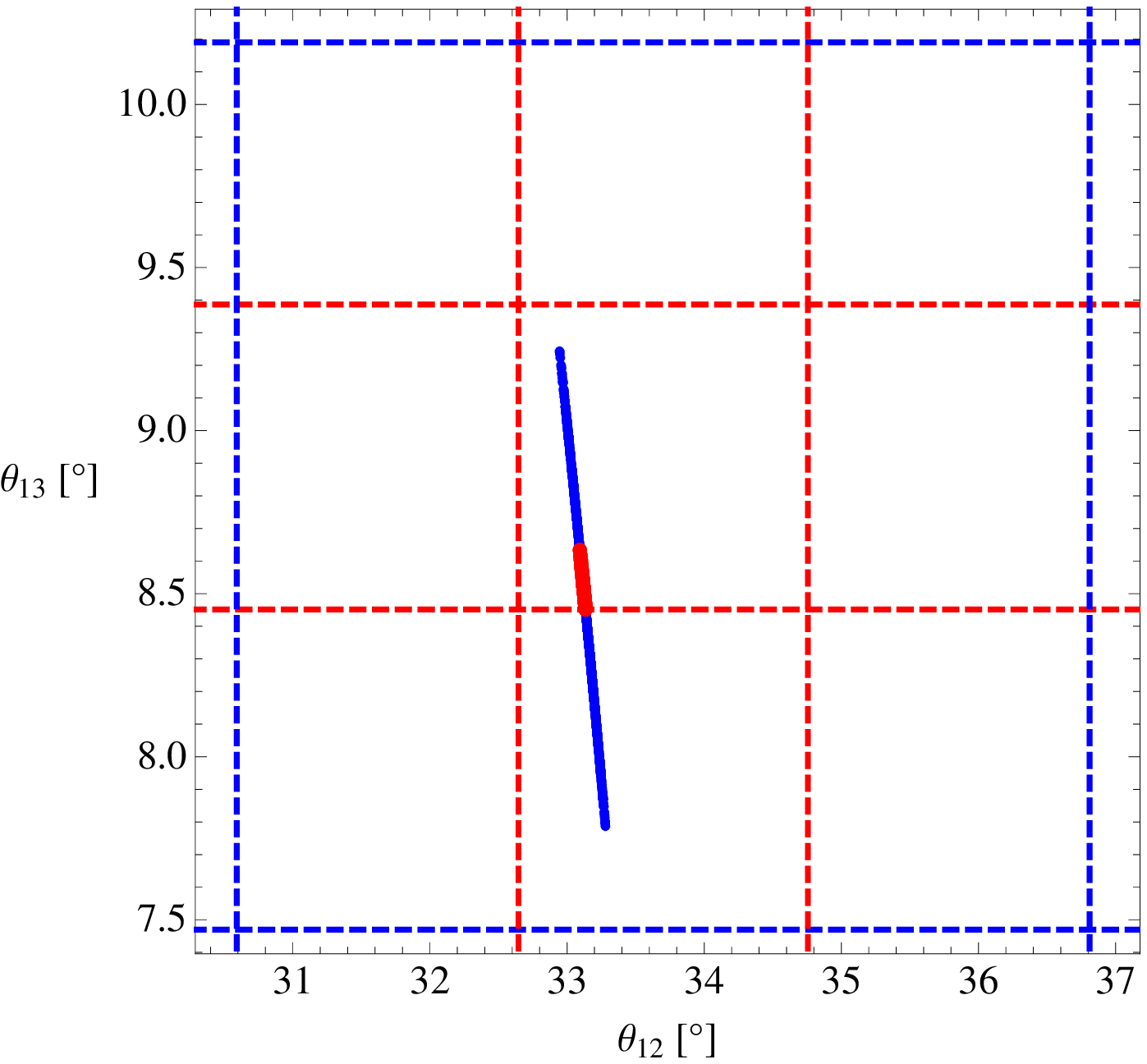}
\includegraphics[scale=0.45]{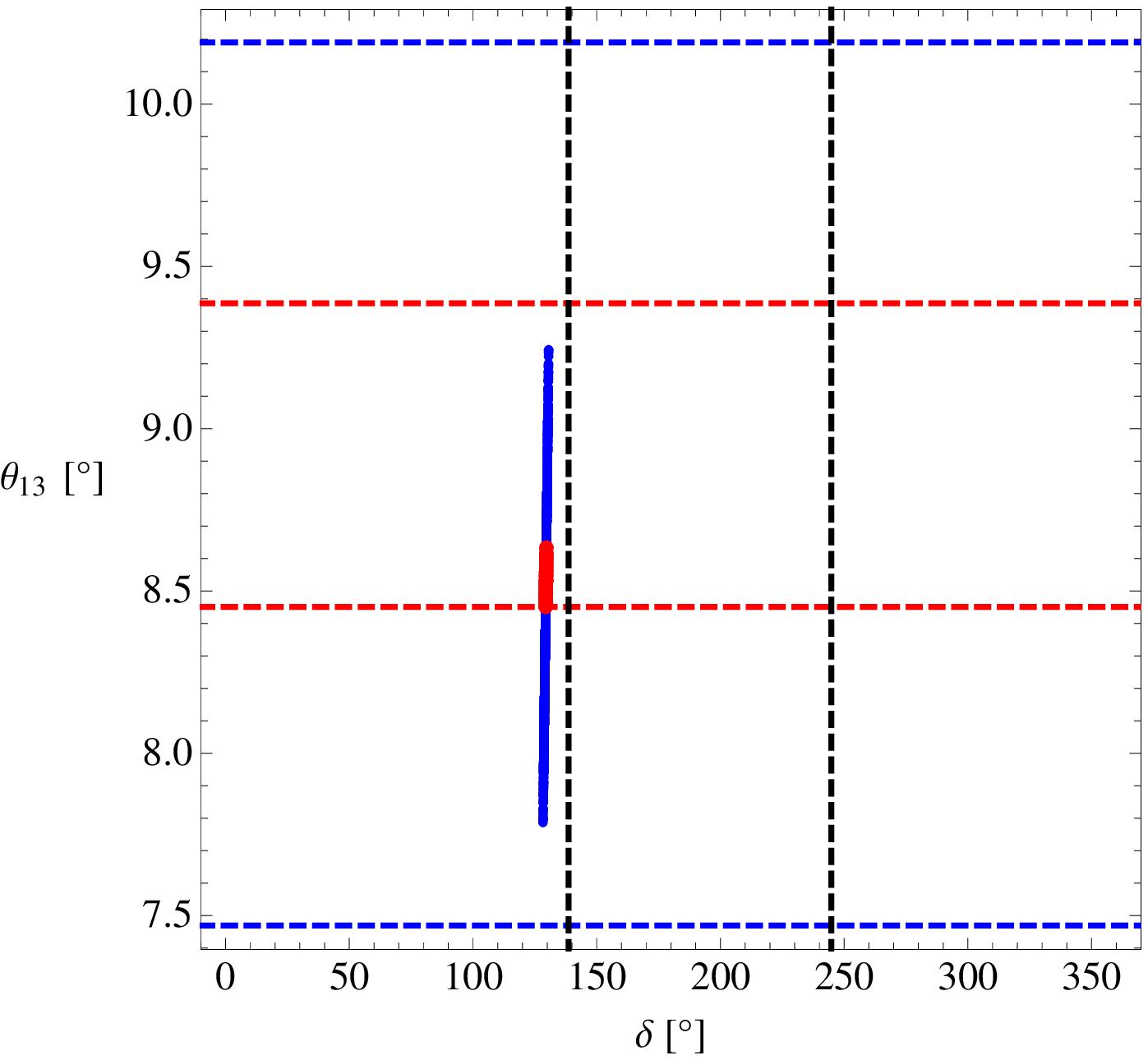}
\includegraphics[scale=0.45]{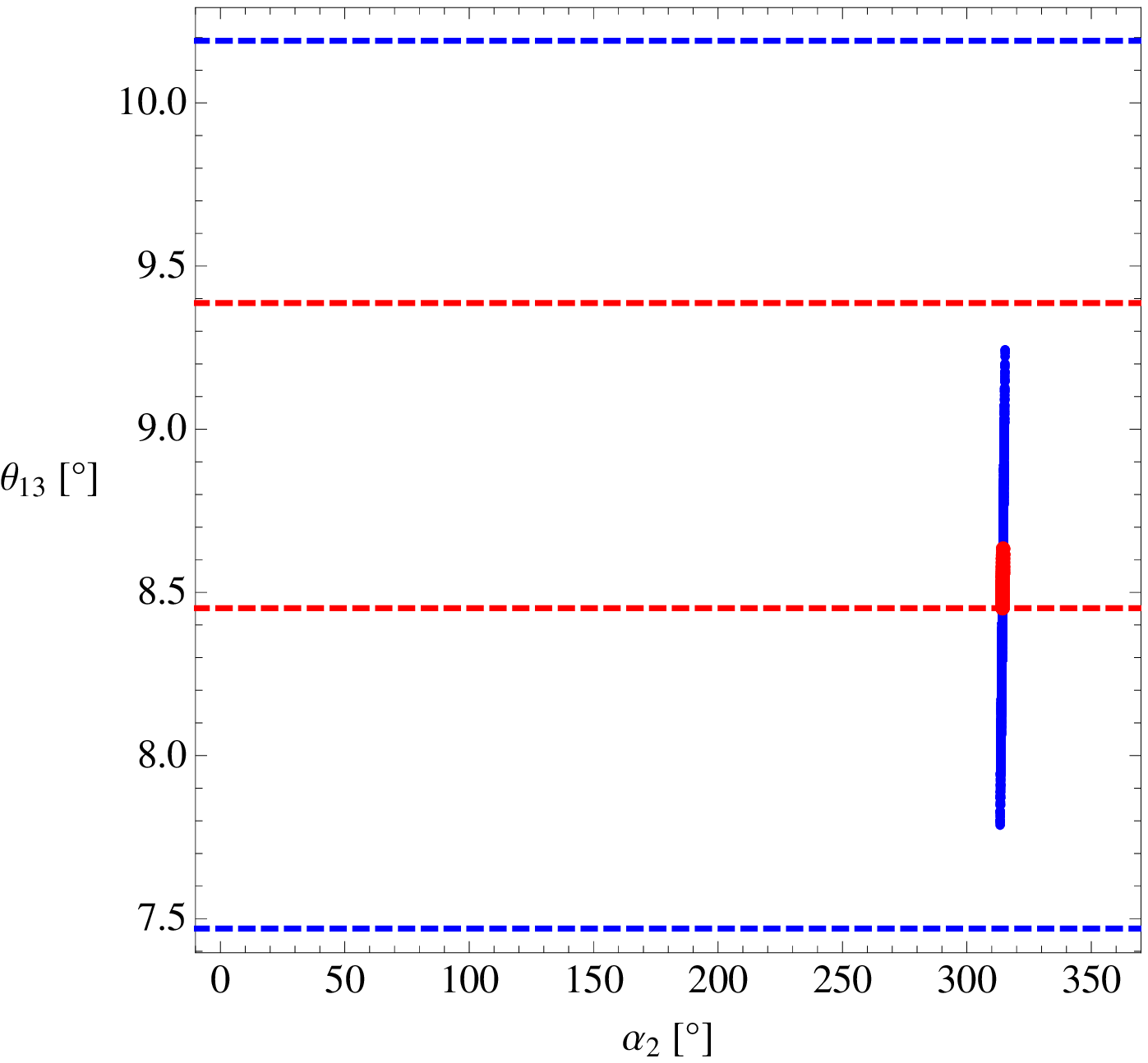}
\includegraphics[scale=0.45]{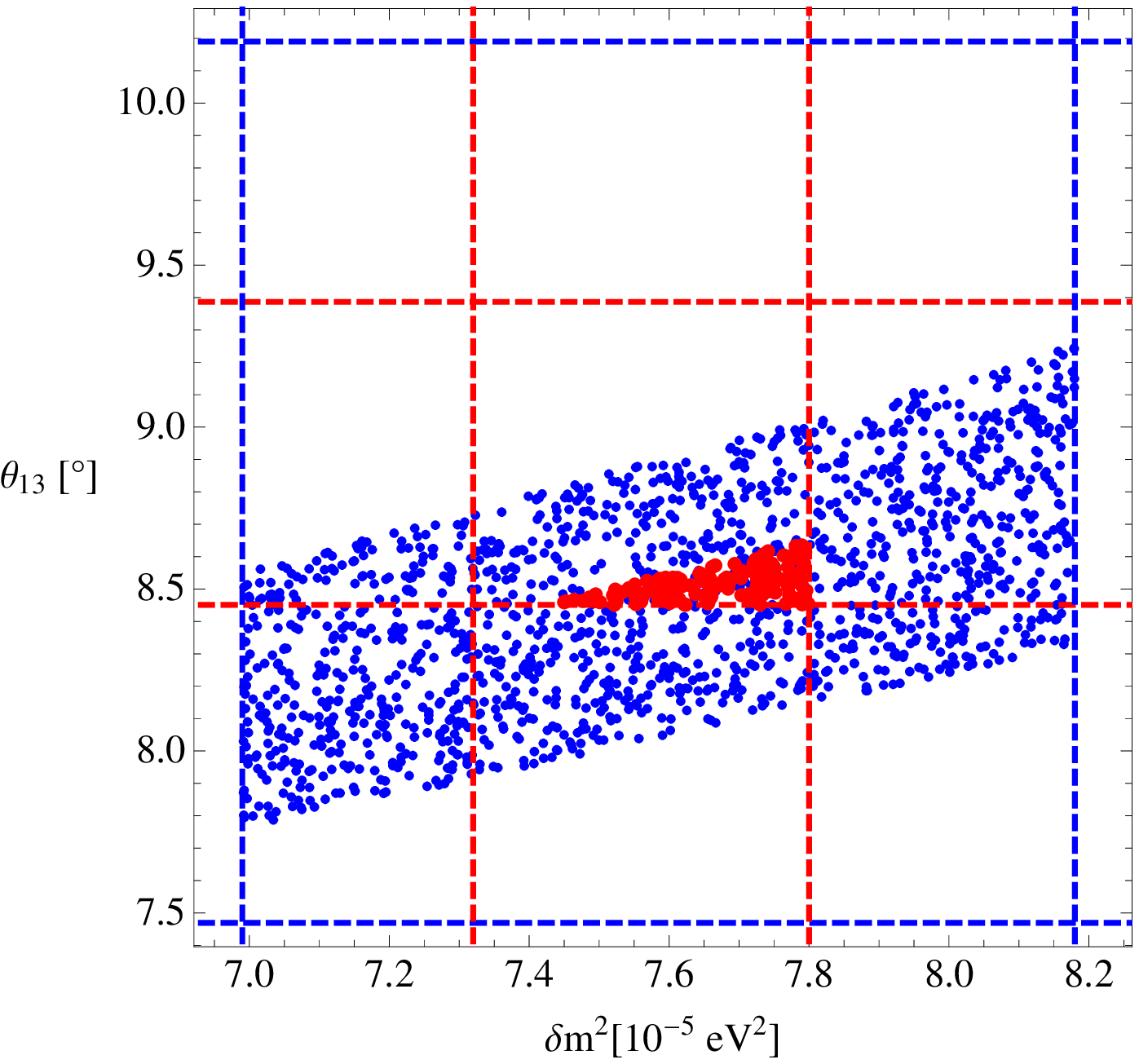}
\includegraphics[scale=0.45]{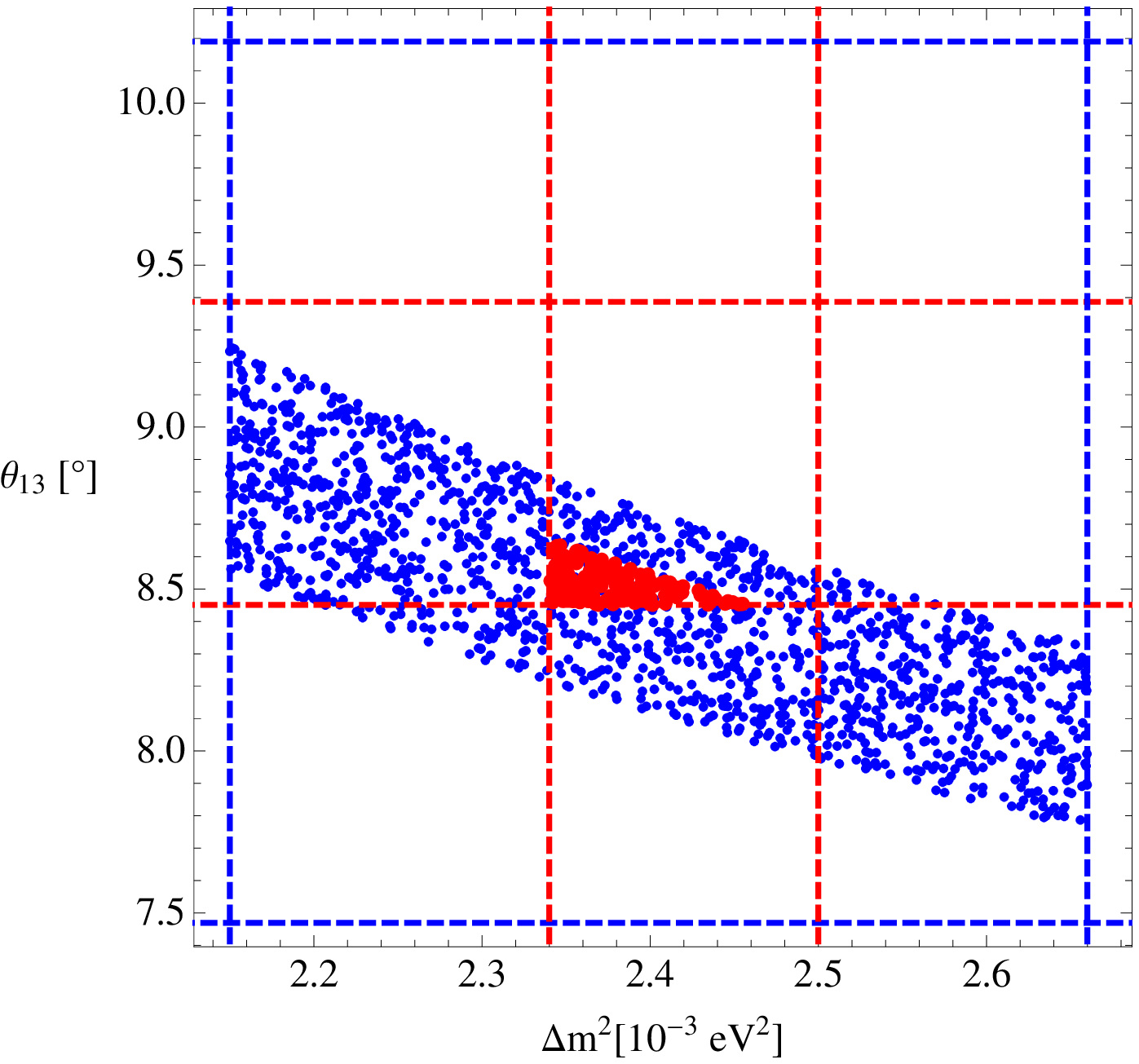}
\caption{The correlations between $\theta_{13}$ and the other two mixing angles
  and the two physical phases in PCSD2. The regions compatible with the 1$\sigma$
  (3$\sigma$) ranges of the mass squared differences and the mixing angles, taken
  from \cite{Fogli:2012ua}, are depicted by the red (blue) points and delimited by
  dashed lines in corresponding colours. The 1$\sigma$ region for the Dirac CP phase
  is shown in black.
  \label{Fig:Correlations1}}
\end{figure}

We turn our discussion now to the neutrino sector.
Here we did not fit the parameters to the observables
because here we are more interested in the allowed
ranges and correlations between different observables
which help in distinguishing this model from other
models.

The effective neutrino mass matrix from
eq.~\eqref{eq:numassmatrix} depends on three
parameters. The neutrino mass scale
$m_a$ the perturbation parameter $\epsilon$
and the relative phase $\alpha$.
The phase $\alpha$ in our model is $\pi/3$
as discussed in section \ref{Sec:Strategy}.
Hence, only two real parameters $m_a$ and
$\epsilon$ completely determine all observables
in the neutrino sector.

We have varied these two parameters randomly and the
results are shown in Figure~\ref{Fig:Correlations1}
where we have used as constraint the fit results of
the Bari group \cite{Fogli:2012ua}. The blue dots
agree with all experimental data within 3$\sigma$
while the red dots agree even within 1$\sigma$.
The dashed lines in the plots label the
corresponding allowed ranges of the observables
on the axes. The 1$\sigma$ range of the leptonic
Dirac phase $\delta$ is
shown in black because it is not measured directly
and the fit results should be taken with a grain
of salt. In the scan we also did not include it
as a constraint.

We are everywhere in good agreement with the experimental
data and we find clear correlations. Especially,
noteworthy is the value for $\theta_{23}$ which lies
around $38.5^\circ$. 
We also make precise
predictions for the CP violating phases. 
One of the
Majorana phases is unphysical because one neutrino
remains massless. The Dirac CP phase has a value of
$\delta \approx 130^\circ$ and the physical Majorana phase is $\alpha_2 \approx
315^\circ$. The Jarlskog determinant $J_\text{CP}$
is around 0.025 and the effective neutrino mass
for neutrinoless double beta decay $m_{ee}$ is of the
order of $3 \times 10^{-3}$~eV, beyond the reach of
current experiments.

\section{Summary}
\label{Sec:Summary}

We have constructed a unified $A_4 \times SU(5)$ model featuring the new type of constrained sequential dominance CSD2 proposed recently in \cite{Antusch:2011ic}.
The $A_4 \times SU(5)$ model, with the CSD2 vacuum alignments $(0,1,1)^T$ and $(1,0,2)^T$,
provides an excellent fit to the present data on quark and lepton masses and mixings, including the measured value of the leptonic mixing angle $\theta_{13}$ from Daya Bay and RENO, with testable predictions for the yet unknown parameters of the leptonic mixing matrix. 

The main idea of the present model is that, with a strong normal hierarchical spectrum (with $m^{\nu}_1 = 0$ by construction since there are only two right-handed neutrinos) the 1-3 angle in the neutrino sector, $\theta_{13}^\nu$, is related to a ratio of neutrino masses by $\theta_{13}^\nu = \frac{\sqrt{2}}{3}\frac{m^\nu_2}{m^\nu_3}$, leading to $\theta^\nu_{13} \sim 5^\circ - 6^\circ$. In addition, the reactor angle receives another contribution from mixing in the charged lepton sector.
The charged lepton mixing induces a correction to $\theta_{13}$ of $\sim 3^\circ$ which adds up constructively with $\theta^\nu_{13}$ to give
\begin{equation}
\theta_{13} \sim 8^\circ - 9^\circ \;, 
\end{equation}
within the range of the measured value from Daya Bay and RENO. 
The constructive addition of the neutrino and charged lepton mixing angles is achieved by 
assuming high energy CP invariance which is spontaneously broken by flavon fields whose
phases are controlled using Abelian $\mathbb{Z}_3$ and $\mathbb{Z}_4$ symmetries as proposed in 
\cite{Antusch:2011sx}. We emphasise that  in our approach
one can either use a ``simple'' CP symmetry, under which the components of the scalar fields transform trivially as 
$\phi_i \rightarrow \phi_i^{*}$, or a ``generalised'' CP symmetry (see e.g.\ \cite{Feruglio:2012cw} and 
references therein) where, in our basis, the triplet fields would transform as $\phi_i \rightarrow U_3 \phi_i^{*}$, with $U_3$ interchanging the second and third component of a triplet representation.

The resulting unified flavour model is highly predictive, as described in section \ref{sec:fit}, since only two parameters determine the neutrino mass matrix, while the charged lepton corrections are fixed by the GUT framework:
In particular, for the Dirac CP phase $\delta$, for the one physical Majorana CP phase $\alpha_2$ and for the atmospheric angle $\theta_{23}$ we obtain the predictions
\begin{equation}
\delta \approx 130^\circ \; , \quad  \alpha_2 \approx 315^\circ \quad   \mbox{and} \quad \theta_{23} \approx 38.5^\circ \;.
\end{equation}
The predictions for $\delta$ and $\theta_{23}$ will be tested by the ongoing and future neutrino oscillation experiments. In addition, for $\theta_{12}$, we predict a value of 
\begin{equation}
\theta_{12} \sim 33^\circ \;, 
\end{equation}
which is slightly smaller than the tribimaximal mixing value but may be tested by a future reactor experiment with $\sim 60$ km baseline, which could measure $\theta_{12}$ with much improved precision \cite{Minakata:2004jt}. Furthermore, in the quark sector, we obtain a right-angled unitarity triangle (with $\alpha \approx 90^\circ$) from the same vacuum alignment techniques for the phases \cite{Antusch:2011sx}, realizing the phase sum rule of \cite{Antusch:2009hq}.

In summary, we have presented a highly predictive new unified model for fermion masses and mixing, which, in fact, represents the first unified indirect family symmetry model in the literature that has been constructed to date that is consistent with all experimental data on quark and lepton mass and mixing angles, and makes definite predictions for CP phases in both the quark and lepton sectors.

\section*{Acknowledgements}
We thank Michael A.~Schmidt and Martin Holthausen for useful discussions about $A_4$ and generalised CP transformations and Christoph Luhn for useful discussions during the early stages of the project.
S.A.\ acknowledges support by the Swiss National Science Foundation,
S.F.K. from the STFC Consolidated ST/J000396/1 and
M.S.\ by the ERC Advanced Grant no. 267985 ``DaMESyFla''.
S.F.K. and M.S. also acknowledge partial support from the EU Marie Curie ITN ``UNILHC'' (PITN-GA-2009-237920) and all authors were partially supported by the European Union under FP7 ITN INVISIBLES (Marie Curie Actions, PITN-GA-2011-289442).

\appendix

\section{Conventions and Notations}
\label{App:Conventions}

In this section we want to summarize briefly our conventions
and define some notation used throughout the main text. We will
follow mainly the notation of \cite{Antusch:2008yc}. The only
difference is a sign in the Majorana phases.

The Yukawa couplings follow the left-right convention
\begin{equation}
 \mathcal{L}_{\text{Yuk}} = - Y_{ij} \overline{\psi_L^i} \psi_R^j H + H.c. \,,
\end{equation}
and for the effective light neutrino mass matrix we use the convention
\begin{equation}
\mathcal{L}_\nu = - \frac{1}{2} \bar L_i (M_\nu)_{ij} L^c_j + H.c. \,,
\end{equation}
where $L$ is the lepton doublet.

In the quark sector we define the CKM matrix by
\begin{equation}
U_{\text{CKM}} = U_{u} U_{d}^\dagger = R_{23} U_{13} R_{12} \;,
\end{equation}
where $U_{u}$ ($U_{d}$) is a unitary matrix diagonalising
$Y_u Y_u^\dagger$ ($Y_d Y_d^\dagger$) and
\begin{equation}
U_{12} = \begin{pmatrix}
  c_{12} & s_{12} \text{e}^{-\ci\delta_{12}} & 0\\
  -s_{12} \text{e}^{\ci \delta_{12}}&c_{12} & 0\\
  0&0&1 \end{pmatrix} \;,
\end{equation}
and similar for $U_{23}$ and $U_{13}$. We use
$c_{12}$ and $s_{12}$ as abbreviations for
$\cos \theta_{12}$ and $\sin \theta_{12}$. The matrices
$R_{23}$ and $R_{12}$ are $U_{23}$ and $U_{12}$
with the complex phases set to zero. In this case
$\delta_{13}$ coincides with the CKM phase
$\delta_{\text{CKM}}$.

For the PMNS matrix we use
\begin{equation}
 U_{\text{PMNS}} = U_{e} U_{\nu}^\dagger = R_{23} U_{13} R_{12} \text{ diag}(\text{e}^{-\ci \alpha_1/2},\text{e}^{-\ci \alpha_2/2},1) \;,
\end{equation}
where the neutrino mass matrix is diagonalized via
\begin{equation}
U_\nu M_\nu M_\nu^\dagger U_\nu^\dagger = \text{diag}(m_1^2,m_2^2,m_3^2) \;.
\end{equation}
and $U_\nu^\dagger = U^{\nu}_{23} U^{\nu}_{13} U^{\nu}_{12}$ 
(note the Hermitian conjugation). This conventions imply a complex conjugation
of the neutrino mass matrix $M_\nu$
compared to our previous CSD2 paper \cite{Antusch:2011ic}
and also the sign of the Majorana phases here is different.

\section{The Renormalizable Superpotential}
\label{App:Messenger}

\begin{table}
\centering
\begin{tabular}{c ccc ccccccc}
\toprule
 & $SU(5)$ & $A_4$ & $U(1)_R$ & $\mathbb{Z}_4$ & $\mathbb{Z}_4$ & $\mathbb{Z}_3$ & $\mathbb{Z}_3$ & $\mathbb{Z}_3$ & $\mathbb{Z}_3$ \\ 
 \midrule 
$\Sigma_1$, $\bar{\Sigma}_1$  &  $\mathbf{5}$, $\bar{\mathbf{5}}$ &  $\mathbf{1}$, $\mathbf{1}$ &  1, 1  &  2, 2  &  0, 0  &  1, 2  &  2, 1  &  2, 1  &  0, 0\\
$\Sigma_2$, $\bar{\Sigma}_2$  &  $\mathbf{5}$, $\bar{\mathbf{5}}$ &  $\mathbf{1}$, $\mathbf{1}$ &  1, 1  &  0, 0  &  0, 0  &  2, 1  &  0, 0  &  1, 2  &  0, 0\\
$\Sigma_3$, $\bar{\Sigma}_3$  &  $\mathbf{5}$, $\bar{\mathbf{5}}$ &  $\mathbf{1}$, $\mathbf{1}$ &  1, 1  &  3, 1  &  3, 1  &  0, 0  &  0, 0  &  2, 1  &  0, 0\\
\midrule
$\Upsilon_1$, $\bar{\Upsilon}_1$  &  $\mathbf{10}$, $\overline{\mathbf{10}}$ &  $\mathbf{3}$, $\mathbf{3}$ &  1, 1  &  1,3  &  3, 1  &  1, 2  &  2, 1  &  2, 1  &  1, 2\\
$\Upsilon_2$, $\bar{\Upsilon}_2$  &  $\mathbf{10}$, $\overline{\mathbf{10}}$ &  $\mathbf{3}$, $\mathbf{3}$ &  1, 1  &  3, 1  &  3, 1  &  0, 0  &  1, 2  &  0, 0  &  1, 2\\
$\Upsilon_3$, $\bar{\Upsilon}_3$  &  $\mathbf{10}$, $\overline{\mathbf{10}}$ &  $\mathbf{3}$, $\mathbf{3}$ &  1, 1  &  3, 1  &  2, 2  &  1, 2  &  1, 2  &  0, 0  &  0, 0\\
\midrule
$\Xi_1$, $\bar{\Xi}_1$  &  $\mathbf{5}$, $\bar{\mathbf{5}}$ &  $\mathbf{3}$, $\mathbf{3}$ &  1, 1  &  3, 1  &  3, 1  &  1, 2  &  1, 2  &  0, 0  &  0, 0\\
$\Xi_2$, $\bar{\Xi}_2$  &  $\mathbf{5}$, $\bar{\mathbf{5}}$ &  $\mathbf{3}$, $\mathbf{3}$ &  1, 1  &  3, 1  &  3, 1  &  0, 0  &  1, 2  &  0, 0  &  1, 2\\
$\Xi_3$, $\bar{\Xi}_3$  &  $\mathbf{5}$, $\bar{\mathbf{5}}$ &  $\mathbf{3}$, $\mathbf{3}$ &  1, 1  &  3, 1  &  2, 2  &  1, 2  &  1, 2  &  0, 0  &  0, 0\\
\midrule
$\Omega_1$, $\bar{\Omega}_1$  &  $\mathbf{10}$, $\overline{\mathbf{10}}$ &  $\mathbf{1}$, $\mathbf{1}$ &  1, 1  &  3, 1  &  3, 1  &  1, 2  &  2, 1  &  2, 1  &  0, 0\\
$\Omega_2$, $\bar{\Omega}_2$  &  $\mathbf{10}$, $\overline{\mathbf{10}}$ &  $\mathbf{1}$, $\mathbf{1}$ &  1, 1  &  3, 1  &  3, 1  &  1, 2  &  0, 0  &  1, 2  &  0, 0\\
$\Omega_3$, $\bar{\Omega}_3$  &  $\mathbf{10}$, $\overline{\mathbf{10}}$ &  $\mathbf{1}$, $\mathbf{1}$ &  1, 1  &  3, 1  &  3, 1  &  2, 1  &  2, 1  &  1, 2  &  0, 0\\
\midrule
$\Gamma_1$, $\bar{\Gamma}_1$  &  $\mathbf{1}$, $\mathbf{1}$ &  $\mathbf{1}$, $\mathbf{1}$ &  0, 2  &  0, 0  &  0, 0  &  2, 1  &  0, 0  &  0, 0  &  1, 2\\
$\Gamma_2$, $\bar{\Gamma}_2$  &  $\mathbf{1}$, $\mathbf{1}$ &  $\mathbf{1}$, $\mathbf{1}$ &  0, 2  &  0, 0  &  0, 0  &  2, 1  &  0, 0  &  2, 1  &  2, 1\\
$\Gamma_3$, $\bar{\Gamma}_3$  &  $\mathbf{1}$, $\mathbf{1}$ &  $\mathbf{1}$, $\mathbf{1}$ &  0, 2  &  0, 0  &  0, 0  &  1, 2  &  2, 1  &  0, 0  &  2, 1\\
$\Gamma_4$, $\bar{\Gamma}_4$  &  $\mathbf{1}$, $\mathbf{1}$ &  $\mathbf{1}$, $\mathbf{1}$ &  0, 2  &  0, 0  &  0, 0  &  2, 1  &  1, 2  &  0, 0  &  0, 0\\
$\Gamma_5$, $\bar{\Gamma}_5$  &  $\mathbf{1}$, $\mathbf{1}$ &  $\mathbf{1}$, $\mathbf{1}$ &  0, 2  &  0, 0  &  0, 0  &  1, 2  &  0, 0  &  1, 2  &  0, 0\\
$\Gamma_6$, $\bar{\Gamma}_6$  &  $\mathbf{1}$, $\mathbf{1}$ &  $\mathbf{3}$, $\mathbf{3}$ &  0, 2  &  0, 0  &  0, 0  &  1, 2  &  2, 1  &  2, 1  &  0, 0\\
$\Gamma_7$, $\bar{\Gamma}_7$  &  $\mathbf{1}$, $\mathbf{1}$ &  $\mathbf{1}$, $\mathbf{1}$ &  0, 2  &  2, 2  &  2, 2  &  0, 0  &  0, 0  &  0, 0  &  0, 0\\
$\Gamma_8$, $\bar{\Gamma}_8$  &  $\mathbf{1}$, $\mathbf{1}$ &  $\mathbf{1}$, $\mathbf{1}$ &  0, 2  &  0, 0  &  2, 2  &  0, 0  &  0, 0  &  0, 0  &  0, 0\\
$\Gamma_9$, $\bar{\Gamma}_9$  &  $\mathbf{1}$, $\mathbf{1}$ &  $\mathbf{1}$, $\mathbf{1}$ &  0, 2  &  0, 0  &  0, 0  &  0, 0  &  1, 2  &  2, 1  &  0, 0\\
\bottomrule
\end{tabular}
\caption{\label{Tab:MessengerFields}
The messenger field content of our model. Every line represents a messenger pair which receives a mass larger than the GUT scale and no cross terms are allowed. In the main text we labelled the messenger mass scale generically with $\Lambda$.}
\end{table}

In this appendix we discuss the full renormalizable superpotential
including the messenger fields which after being integrated out
give the effective operators as discussed before.

We start with the superpotential bilinear in the fields which
is in our case only the mass terms for the messengers
\begin{equation}
 \mathcal{W}^{\text{ren}}_\Lambda = M_{\Sigma_i} \Sigma_i \bar \Sigma_i + M_{\Upsilon_i} \Upsilon_i \bar \Upsilon_i + M_{\Xi_i} \Xi_i \bar \Xi_i + M_{\Omega_i} \Omega_i \bar \Omega_i + M_{\Gamma_i} \Gamma_i \bar \Gamma_i \;.
\end{equation}
The full list of messenger fields is given in
Table~\ref{Tab:MessengerFields} where every line
is a messenger pair which receives a mass larger
than the GUT scale so that they can be integrated
out to give the desired effective operators. To
simplify the notation before we have introduced
the messenger scale $\Lambda$ as shorthand which
is related to the individual messenger masses
with order one coefficients.

Note that in the superpotential bilinear in the
fields no $\mu$-term for the Higgs fields appears.
This term is forbidden by symmetries and in
combination with a NMSSM like mechanism helps
to increase the Higgs mass to the experimentally
determined value. A possible singlet field $S$ with couplings
$S (H_5 \bar H_5 + H_{45} \bar H_{45})$ would not
appear anywhere else in the superpotential with the
symmetries and field content as specified in Tables~
\ref{Tab:Matter+HiggsFields}, \ref{Tab:FlavonFields},
\ref{Tab:DrivingFields} and \ref{Tab:MessengerFields}.

\begin{figure}
\centering
\includegraphics[width=\textwidth]{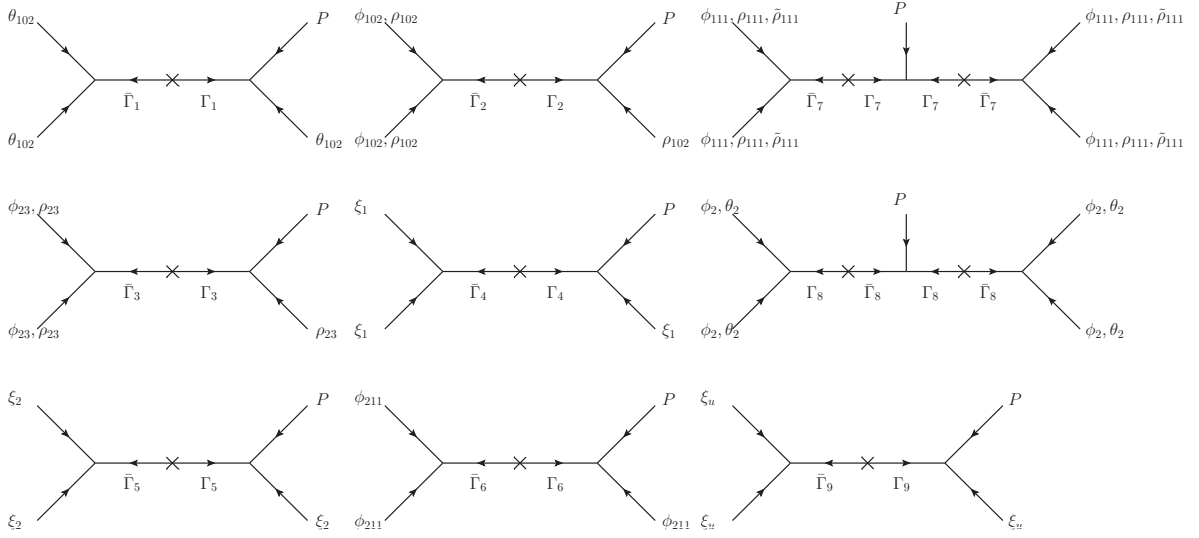}
\caption{
The supergraphs before integrating out the messengers for the flavon sector (only diagrams are shown which give non-renormalizable contributions).
\label{Fig:FlavonMessenger} }
\end{figure}

The next step in our discussion of the renormalizable
superpotential is the flavon sector. The full potential
for this sector reads (dropping for the sake of simplicity
order one coefficients)
\begin{align}
  \mathcal{W}^{\text{ren}}_{\text{flavon}} &= O_{1;2} \phi_1  \phi_2 + O_{1;3} \phi_1  \phi_3 + O_{2;3} \phi_2  \phi_3 + O_{111;211} \phi_{111}  \phi_{211} + O_{111;23}  \phi_{111}  \phi_{23} \nonumber\\
& + O_{23;211} \phi_{23}  \phi_{211} + O_{2;102} \phi_2  \phi_{102} + O_{211;102} \phi_{211}  \phi_{102} + O_{1;23}  \phi_1  \phi_{23} \nonumber\\
& + A_1  \phi_1  \phi_1 + A_2  \phi_2  \phi_2 + A_3 \phi_3  \phi_3 + A_{111} \left( \phi_{111}^2  + \phi_{111}  \rho_{111}+ \tilde \phi_{111} \rho_{111}\right) \nonumber\\
& + P \Gamma_9 \xi_u + \bar \Gamma_9 \xi_u^2 + P \Gamma_8^2 + \bar \Gamma_8 \phi_2^2 + \bar \Gamma_8 \theta_2^2 + P \Gamma_7^2 + \bar \Gamma_7 ( \phi_{111}^2 + \rho_{111}^2 + \tilde \rho_{111}^2 ) \nonumber\\
& + P \phi_{211} \Gamma_6 + \phi_{211}^2  \bar \Gamma_6 + P \xi_2 \Gamma_5 + \xi_2^2 \bar \Gamma_5 + P \xi_1 \Gamma_4 + \xi_1^2  \bar \Gamma_4 + P \rho_{23} \Gamma_3 + (\phi_{23}^2 + \rho_{23}^2) \bar \Gamma_3  \nonumber\\
& + P \rho_{102} \Gamma_2 + ( \phi_{102}^2 + \rho_{102}^2) \bar \Gamma_2+ P \theta_{102} \Gamma_1 + \theta_{102}^2 \bar \Gamma_1 \;.
\end{align}
The first three lines of this superpotential have 
already been discussed in the flavon alignment section~\ref{Sec:Flavon}
while the last four lines are needed to fix the phases
of the various flavon vevs. For instance, the
messenger pair $\Gamma_1$ and $\bar \Gamma_1$
gives after integrating out the effective operator
$1/\Lambda P \theta_{102}^3$ where in this
case $\Lambda$ stands for $M_{\Gamma_1}$ multiplied
by real order one couplings. This operator fixes the
phase of $\langle \theta_{102} \rangle$ up to a
discrete choice as discussed before.

We will not list
here all of the effective operators because they
have already appeared in our superpotential
for the flavon alignment and they can also be read
off from the diagrams in Figure~\ref{Fig:FlavonMessenger}
after contracting the messenger propagators to points.

\begin{figure}
\centering
\includegraphics[scale=0.45]{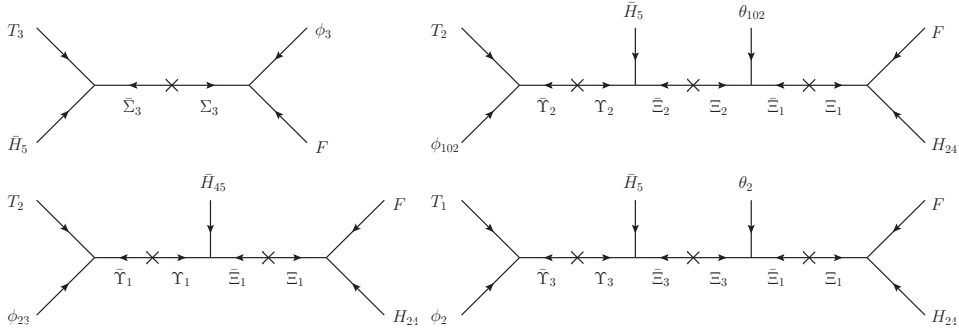}
\caption{
The supergraphs before integrating out the messengers for the down-type quark and charged lepton sector.
\label{Fig:DownMessenger} }
\end{figure}

\begin{figure}
\centering
\includegraphics[scale=0.55]{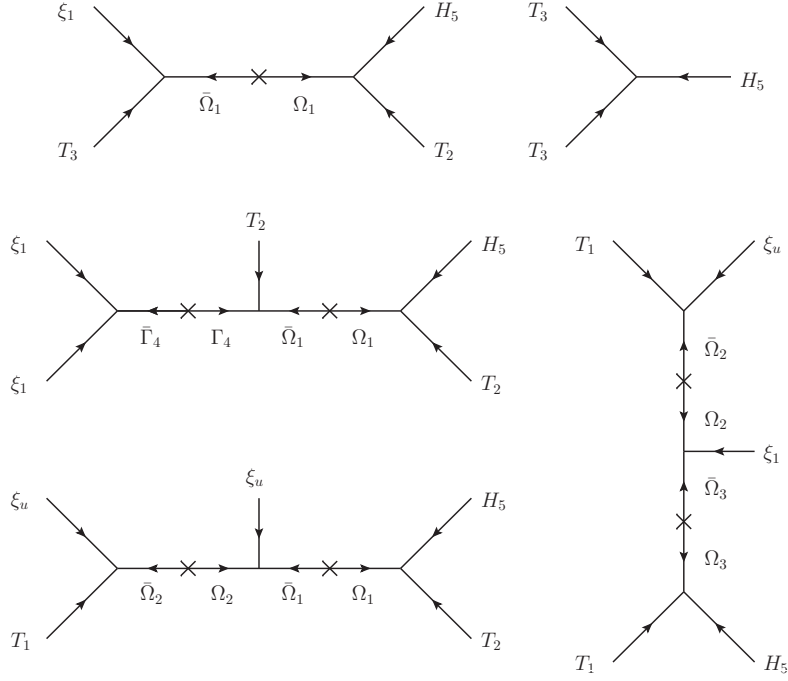}
\caption{ The supergraphs before integrating out the messengers for
the up-type quark sector.
\label{Fig:UpMessenger} }
\end{figure}

\begin{figure}
\centering
\includegraphics[scale=0.6]{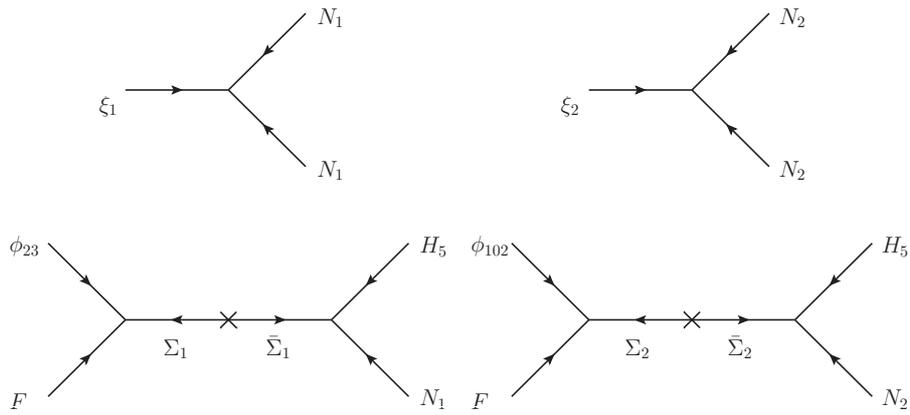}
\caption{
The supergraphs before integrating out the messengers for the neutrino sector.
\label{Fig:NeutrinoMessenger} }
\end{figure}

For the renormalizable couplings including the matter and Higgs
fields we find the renormalizable superpotential (again
dropping order one coefficients)
\begin{align}
 \mathcal{W}^{\text{ren}}_d &= T_3 \bar H_5 \bar \Sigma_3 + F  \phi_3  \Sigma_3 + T_2  \phi_{23} \bar \Upsilon_1 + \bar H_{45} \Upsilon_1 \bar \Xi_1 + F  H_{24}  \Xi_1 + T_2  \phi_{102}  \bar \Upsilon_2 + \bar H_5 \Upsilon_2 \bar \Xi_2 \nonumber\\
 & + \theta_{102} \Xi_2 \bar \Xi_1 + T_1  \phi_2 \bar \Upsilon_3 + \bar H_5 \Upsilon_3 \bar \Xi_3 + \theta_2 \Xi_3  \bar \Xi_1 \;,\\
 \mathcal{W}^{\text{ren}}_u &= T_1  H_5 \Omega_3 + \xi_1 \Omega_2 \bar \Omega_3 + T_1  \xi_u \bar \Omega_2 + \Omega_2 \xi_u \bar \Omega_1 + T_2 \Gamma_4 \bar \Omega_1 \nonumber\\
& + \bar \Gamma_4 \xi_1^2 + T_2  H_5 \Omega_1 + T_3 \xi_1 \bar \Omega_1 + T_3^2  H_5 \;, \\
 \mathcal{W}^{\text{ren}}_\nu &= \xi_1 N_1^2 + \xi_2  N_2^2 + F \phi_{23}  \Sigma_1 + N_1 H_5  \bar \Sigma_1 + F \phi_{102} \Sigma_2 + N_2 H_5 \bar \Sigma_2  \;. 
\end{align}
After integrating out the heavy messenger fields
we end up with the non-renormalizable operators
as discussed in section~\ref{Sec:Model},
cf.\ also Figures~\ref{Fig:DownMessenger}-\ref{Fig:NeutrinoMessenger}.

In addition to the renormalizable operators discussed
so far there are six more operators allowed by the
symmetries which are
\begin{equation}
 \mathcal{W}^{\text{ren}}_{\text{neg}} =
T_1 \Gamma_9 \bar \Omega_1 + T_2 \Gamma_9 \bar \Omega_3 + \Xi_1 \bar \Xi_3 \phi_2 + \Gamma_9 \Omega_1 \bar \Omega_2 + \Gamma_4 \bar \Omega_2 \Omega_3 + \Gamma_1 \Xi_1 \bar \Xi_2 \;.
\end{equation}
The first two operators contribute effectively to the
$T_1 T_2 H_5 \xi_u^2$ operator already present and for the sake of
simplicity we have not shown them in Figure~\ref{Fig:UpMessenger}.
The third operator
generates the dimension six operator $F T_2 \bar H_{45}
H_{24} \phi_2  \phi_{23}$ which gives a contribution
to the 2-2 element of the down-type quark and charged
lepton Yukawa matrix. In fact the correction has the
same phase and the same $SU(5)$ Clebsch--Gordan
coefficient as the leading order coefficient so that
we can safely neglect it.
The last three operators finally give, after integrating out the
heavy messengers, dimension seven
and eight operators which give
only small corrections (in our model we have discussed
operators up to dimension six). The dimension seven
operators, for instance, are induced by $\Gamma_9
\Omega_1 \bar \Omega_2$ which gives corrections to the $1-3$
and $2-3$ elements of the up-type quark Yukawa matrix
which are very small compared to all other elements which
are generated at maximum by a dimension five operator.

\end{document}